\documentclass[twocolumn]{aastex62}
\pdfoutput=1 
\usepackage{amsmath,amstext}
\usepackage[T1]{fontenc}
\usepackage[figure,figure*]{hypcap}

\shorttitle{EUV fluxes of Very Low-Mass Stars}
\shortauthors{J. J. Drake et al.}

\begin{document}


\title{Pointing {\it Chandra} Toward the Extreme Ultraviolet Fluxes of Very Low-Mass Stars}

\author[0000-0002-0210-2276]{Jeremy J.\ Drake}
\author[0000-0002-3869-7996]{Vinay L.\ Kashyap}
\author[0000-0002-2096-9586]{Bradford J.\ Wargelin}
\author[0000-0002-0826-9261]{Scott J.\ Wolk}
\affiliation{Smithsonian Astrophysical Observatory, 60 Garden Street, Cambridge MA02138, USA}

\begin{abstract}
The X-ray and EUV emission of stars plays a key role in the loss and evolution of the atmospheres of their planets. The coronae of dwarf stars later than M6 appear to behave differently to those of earlier spectral types and are more X-ray dim and radio bright.   Too faint to have been observed by the Extreme Ultraviolet Explorer,  their EUV behavior is currently highly uncertain.  We have devised a method to use the {\it Chandra} X-ray Observatory High Resolution Camera to provide a measure of EUV emission in the 50-170~\AA\ range and have applied it to the M6.5 dwarf LHS~248 in a pilot 10~ks exposure.  Analysis with model spectra using simple, idealised coronal emission measure distributions inspired by an analysis of {\it Chandra} HETG spectra of the M5.5 dwarf Proxima Cen and results from the literature, finds greatest consistency with a very shallow emission measure distribution slope, $DEM \propto T^{3/2}$ or shallower, in the range $\log T=5.5$--$6.5$.  Within  $2\sigma$ confidence, a much wider range of slopes can be accommodated.  Model spectra constrained by this method can provide accurate (within a factor of 2--4) synthesis and extrapolation of EUV spectra for wavelengths $<400$--500~\AA.  At longer wavelengths models are uncertain by an order of magnitude or more, and depend on the details of the emission measure distribution at temperatures $\log T < 5.5$.  The method is sensitive to possible incompleteness of plasma radiative loss models in the 30--170~\AA\ range for which re-examination would be warranted.
\end{abstract}

\keywords{Stars: activity, flare, late-type, coronae, low-mass --- Sun: X-rays --- X-rays: stars --- planets and satellites: atmospheres}

\section{Introduction}
\label{s:intro}

Magnetic activity on stars like the Sun with outer convection zones is manifest through energetic photon and particle radiation in the form of UV to X-ray emission from hot coronal plasma and a hot magnetized wind.  
Both photon and particle emissions have a stochastic component where stored magnetic potential energy is more impulsively released in flares and coronal mass ejections.  Magnetic activity is driven by stellar rotation and so gradually declines with time as stars lose angular momentum to stellar winds \citep[e.g.][]{Skumanich:72,Wright.etal:11,Garraffo.etal:18}. Understanding of the physics of stellar coronae remains incomplete, and its study continues to be a cottage industry in modern astrophysics.  

The growing realization that planetary systems are extremely common around Sun-like stars has led to a resurgence in interest in stellar magnetic activity.  Exoplanets have catalyzed a shift in focus from coronal physics itself to the effects of energetic stellar radiation on planetary atmospheres and environments.  
Both UV--X-ray emission and stellar winds are agents of atmospheric destruction.  In our own solar system, the action of the solar wind stripped copious amounts of water from the atmosphere of Mars during the Noachian period about 4 Gyr ago \citep[e.g.][]{Jakosky.etal:15}.  The roles of X-ray and EUV radiation in exoplanet atmospheric evaporation has been highlighted in several recent studies  \citep[e.g.][]{Penz.Micela:08,Murray-Clay.etal:09,Sanz-Forcada.etal:10,Owen.Jackson:12,Owen.Wu:13,Chadney.etal:15,Owen.Wu:16}, and 
there is now convincing evidence that planetary envelopes have been reduced and evaporated by the host star EUV and X-ray emission \citep[e.g.][]{Sanz-Forcada.etal:11,Owen.Wu:13}.  Planets which have received higher integrated doses of EUV and X-ray exposure tend to be smaller objects that appear to have lost part or all of their envelopes.  

Based on a simplifying assumption that EUV and X-ray emission decline with time (due to stellar spin down) at the same rate, \citet{Owen.Jackson:12} find that X-rays dominate mass loss during early times when stellar activity is very high. At later times, EUV emission dominates. 
In their modelling, \citet{Owen.Jackson:12} adopted the assumption of lockstep decline in EUV and X-ray fluxes because, in their words, ``The evolution of the EUV luminosity is still observationally rather unclear''.  Current evidence seems to point to X-ray decline being more rapid than EUV decline \citep[e.g.][]{Sanz-Forcada.etal:11,Chadney.etal:15}, but it is now apparent that a deeper understanding of how stellar EUV and X-ray fluxes relate to each other and how they change with time is key to understanding planetary atmospheric evolution.

The ``Unobservable Extreme Ultraviolet'' is a phrase coined some decades ago representing the largely mistaken belief that the ISM would be so opaque at EUV wavelengths that nothing could be observed.  It is apt today because there is currently no facility capable of observing cosmic (i.e. non-solar) EUV emission. 
While the longer wavelengths of the EUV bandpass (here loosely defined as 100--912~\AA) are indeed essentially inaccessible for all but the most nearby stars, the {\it Extreme Ultraviolet Explorer} ({\it EUVE}) did accumulate  spectra of a number of either nearby or particularly active late-type stars in the 70-750~\AA\ range \citep[e.g.][]{Drake:96,Bowyer.etal:00}.  These data have been used by different authors to examine the EUV flux and how it can be estimated using different proxies, such as X-ray flux \citep{Sanz-Forcada.etal:11} or Ly$\alpha$ emission \citep{Linsky.etal:14,Youngblood.etal:16}.  However, these studies have been limited to stars of spectral type M5 and earlier, for the simple fact that later types were too faint for {\it EUVE} to observe.

The remarkable discoveries of Proxima b \citep{Anglada-Escude.etal:16} and four additional planets around TRAPPIST-1, making a total of seven currently known \citep{Gillon.etal:17}, has highlighted the case for M dwarfs being 
favorable candidate host stars with nearby Earth-like exoplanets amenable to detailed study.  While M dwarfs are the most numerous stars in the Galaxy, and will likely host the largest number of conveniently nearby exoplanets, their magnetic activity represents a proportionally larger fraction of their bolometric output than earlier type stars, and longer spin down timescales means this output stays higher for longer.  Habitable zone planets around M dwarfs are then even more susceptible to EUV, X-ray, and stellar wind destruction. 

The EUV flux from late-type stars originates from the chromosphere, the transition region, and the corona \citep[e.g.][]{Drake:98,Linsky.etal:14}, and it is necessary to have an accurate description of all of these regions to understand the EUV radiative output of stars. 
Late-type M dwarfs (spectral types $>$M5) such as  TRAPPIST-1 (M8 V) present a difficult case for the use of proxies to infer EUV radiation output. 
 The effective temperature of an M8 dwarf is only about 2500~K and its atmosphere is highly neutral. These stars are fully convective and it has been conjectured that their dynamo activity is fundamentally different.  \citet{Wright.Drake:16} have shown that fully convective stars down to type M5.5V do follow the same rotation-activity relation as earlier types.  However, relative to bolometric luminosity both H$\alpha$, $L_{H_\alpha}/L_{bol}$, and X-ray $L_X/L_{bol}$, outputs are seen to decline toward later M spectral types \citep[e.g.][]{Mohanty.Basri:03, Berger.etal:10,Cook.etal:14}, while radio output tends to increase \citep[e.g.,][]{Williams.etal:14}, with changes in behaviour appearing to set in around spectral type M7-M8.

Since late M-type dwarfs do not appear to follow the same X-ray behavior as earlier types, we have no indication that their EUV fluxes do either, or how they behave at all.  In the absence of an EUV observational capability, other observational and theoretical methods that are able to infer the EUV fluxes of late M dwarfs would be of considerable value.  \citet{Peacock.etal:18} have developed theorical atmospheric models that appear promising for constraining emission in the longer wavelengths of the EUV bandpass for very low mass stars. At shorter wavelengths, EUV emission stems from coronal temperatures $\log T \geq 5.5$.  While X-ray observations can be used to develop coronal models for temperatures $\log T \geq 6$, there remains a gap in temperature coverage between a few $10^5$~K and $10^6$~K that is crucial for constraining EUV fluxes.

Here, we develop a method to utilize the {\it Chandra} X-ray Observatory {\it High Resolution Camera} spectroscopic detector (HRC-S) to provide EUV constraints.  This detector has  sensitivity well into the EUV range, out to 170~\AA. We devise an observing method to utilize an off-axis region of the detector with a thinner Al coating that has higher sensitivity at wavelengths longward of 44~\AA\ than the on-axis filter.  We analyse a pilot observation of the late M dwarf LHS~248 and show that the observing method can provide quite tight observational constraints on its EUV flux in the 100-400~\AA\ range.  The observing rationale is explained in Sect.~\ref{s:euv_sens}.  In Sect.~\ref{s:obs} the various observations utilized in the study are described, and their analysis is related in Sect.~\ref{s:anal}.  We discuss the issues involved in interpreting the data in Sect.~\ref{s:disc} and reach our conclusions in Sect.~\ref{s:con}.

\section{EUV Sensitivity of the Chandra HRC-S}
\label{s:euv_sens}

The {\it Chandra} X-ray Observatory \citep{Weisskopf.etal:03} LETG+HRC-S instrument reaches to wavelengths of 170~\AA\ (and even longer with offset pointing), which is well into the EUV part of the spectrum.  However, it is essentially a bright object spectrometer: depending on the intervening interstellar medium absorption, spectra are spread over up to 300\,mm of microchannel plate detector, and for faint targets the signal becomes background dominated and swamped.  With perhaps the exception of our second nearest star, Proxima Cen (M6.5V; see Sect.~\ref{s:proxletg} below), very late fully convective M dwarfs are generally beyond its reach. 

\begin{figure}{}
\center
\includegraphics[width=0.46\textwidth]{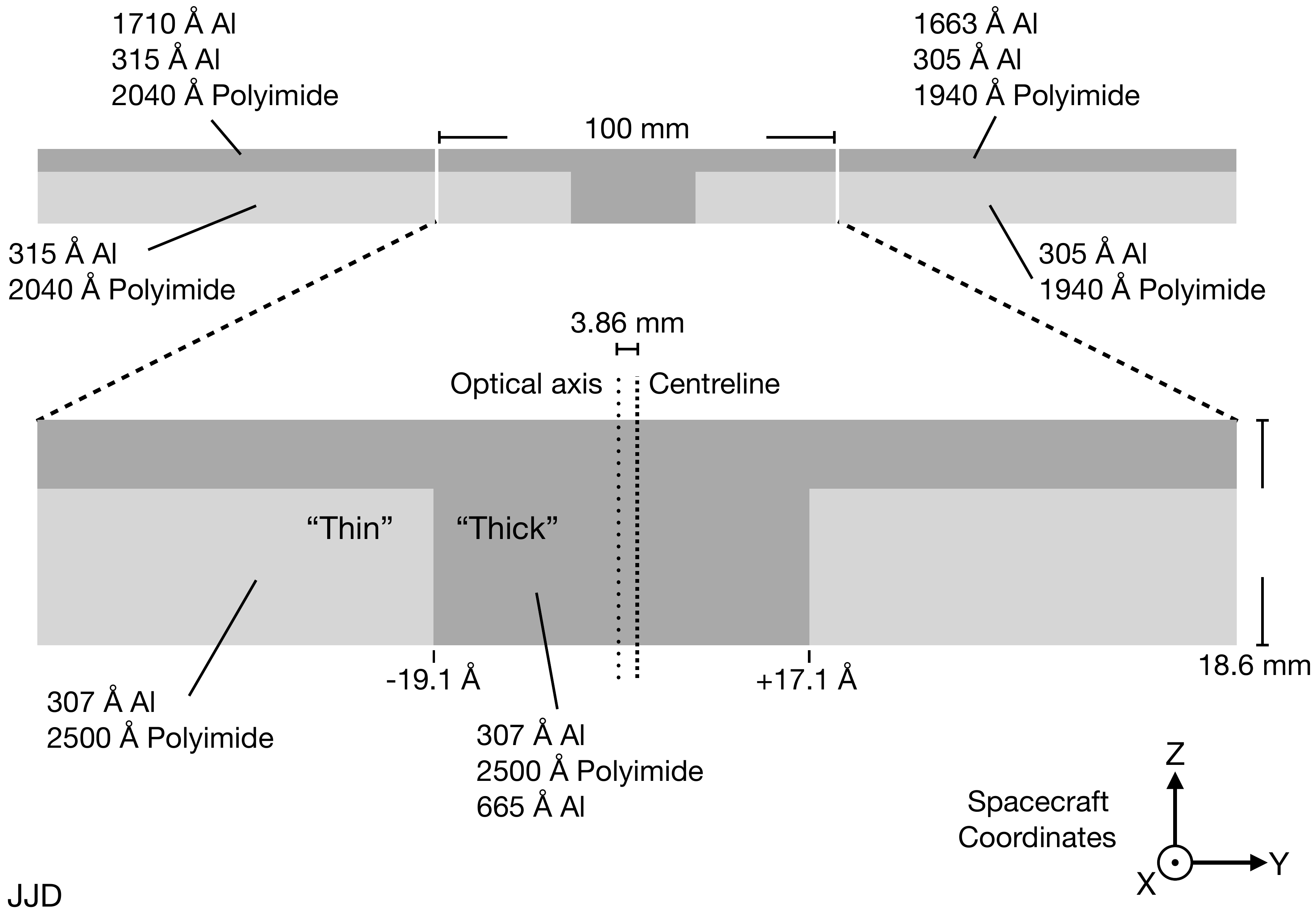} 
\caption{A schematic of the {\it Chandra} HRC-S UV/Ion shield with the centre segment shown enlarged for detail. The filter characteristics of the ``thin'' and ``thick'' Al regions of the central segment referred to and used in the observations described here are noted.  The observation of LHS~248 was obtained by pointing 5.03~arcmin in the positive $y$ direction so as to place the source at the thick/thin Al filter boundary in the $-y$ direction, to the left of centre as shown here.
}
\label{f:hrcs}
\end{figure}

A schematic of the HRC-S UV/Ion shield (UVIS), designed to thwart optical and UV throughput and low-energy protons, is illustrated in Figure~\ref{f:hrcs}.  The UVIS comprises three segments of Al-coated polyimide film.  The two outer segments have the same characteristics within manufacturing tolerances, while the center segment is slightly different.  Each filter segment covers a stack of microchannel plates and a charge readout grid connected to signal amplifiers that make up the rest of the detector.  

{}

The different segments were designed to handle the different wavelengths of the first order dispersed LETG spectrum. Each has a strip of thicker Al coating, serving as a "low energy" suppression filter and designed for using primarily on the higher dispersed LETG orders throughput, and a larger section of thinner Al as the main first order detection region.

The centre segment, together with the underlying microchannel plate detector, also serves as the back-up imaging detector to the HRC-I, as well as being the prime {\it Chandra} instrument for high timing resolution observations.  Consequently, the region at the centre where the aim point lies---the vertical part of the ``T'' shape in Figure~\ref{f:hrcs}---has a thicker Al coating for more aggressive UV and optical attenuation.  Away from the detector centre, during normal operation with the LETG in place,  there is no signal from dispersed UV or optical photons, and the out-of-band rejection requirements are less stringent.  Consequently, outside of the ``T'' region the Al coating is much thinner.   This filter region has significantly larger transmittance than the centre region with the thicker Al coating.  

\begin{figure}{}
\center
\includegraphics[width=0.46\textwidth]{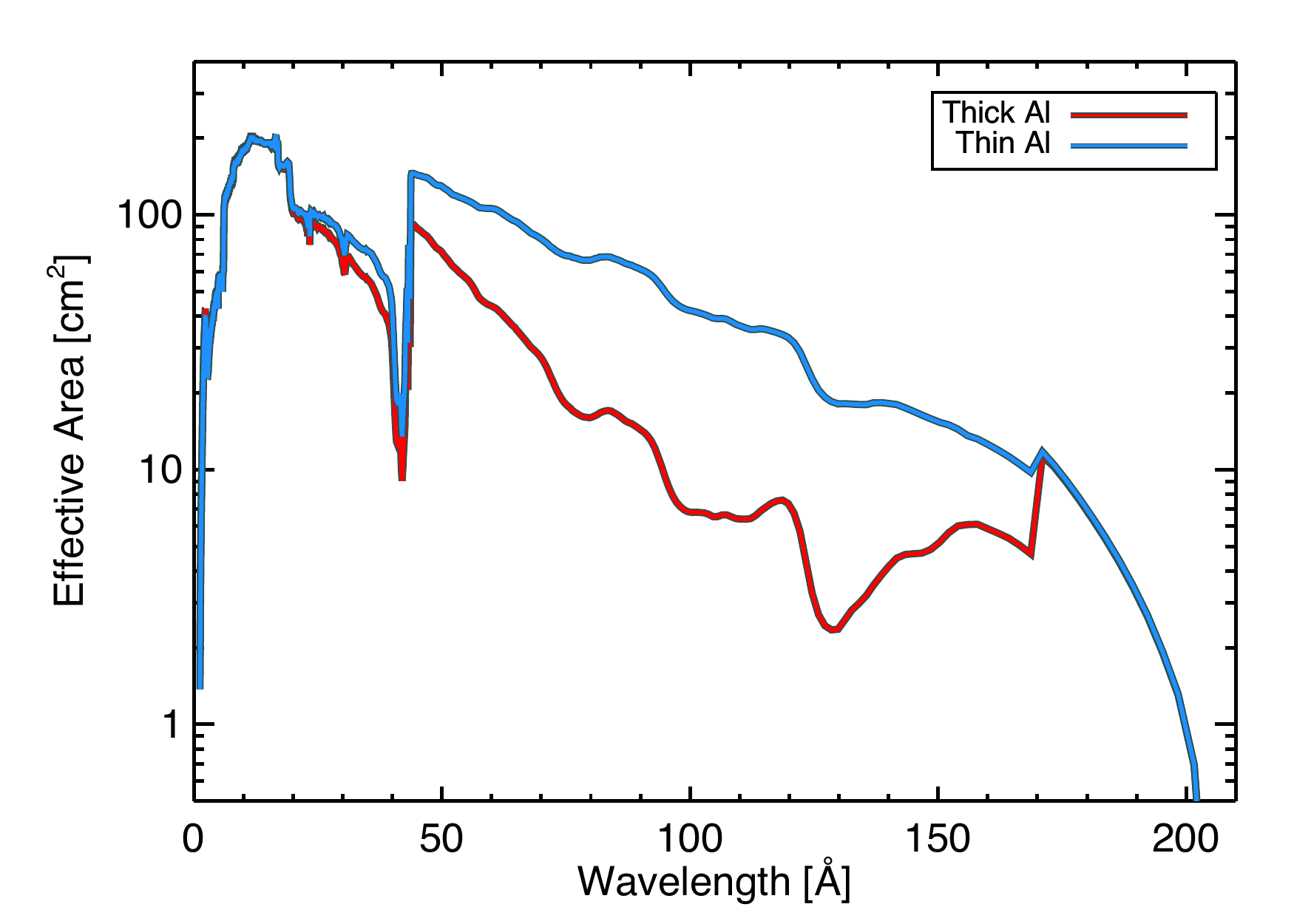} 
\caption{The effective areas of the {\it Chandra} HRC-S ``thin'' and ``thick'' Al filter regions computed at the location of the region boundary in the $-y$ direction from the detector centre.
}
\label{f:eff_areas}
\end{figure}

The effective areas of the HRC-S ``thin'' and ``thick''' Al regions at the filter boundary in the $-y$ direction from the aimpoint are illustrated in Figure~\ref{f:eff_areas}.  The differences become pronounced for wavelengths longer than the C~K edge near 44~\AA, up to the Al L edge near 170~\AA, with the area being larger on the thin filter by typical factors of 2--4. It is this difference in effective areas that gives us some power to discriminate between shorter and longer wavelength emission.

{}

\section{Observations}
\label{s:obs}

Details of the {\it Chandra} observations of LHS~248 and Proxima Centuri analysed in this study are listed in Table~\ref{t:obs}.  All data reduction and computation of instrument response files followed standard procedures using the {\it Chandra} Interactive Analysis of Observations (CIAO; \citealt{Fruscione.etal:06}) software framework.
Pertinent details for each are described further below.

\subsection{LHS 248}

LHS~248 (DX~Cnc, GJ~1111) has been assigned credible spectral types in the range M6~V \citep{Davison.etal:15} to M7.1~V \citep{Terrien.etal:15}, with the most common designation being M6.5~V \citep{Alonso-Floriano.etal:15}.  At a distance of 3.58 parsecs \citep{:18} LHS~248 is the 18th closest star (or star system) to the Sun.  It is a rapid rotator and a magnetically active flare star \citep[e.g.][]{Reiners.Basri:07}, with a period of 0.46~days \citep{Morin.etal:10,Newton.etal:18} and a probable member of the Castor moving group with an age of 200 Myr \citep[][and references therein]{Caballero:10}.  LHS~248 was detected in a 16~ks ROSAT PSPC pointed observation with an X-ray luminosity of $4\times 10^{26}$~erg~s$^{-1}$ in the survey by \citet{Schmitt.etal:95}. The corresponding ratio between X-ray and bolometric luminosities is $\log L_X/L_{bol}=-3.9$ \citep{Cook.etal:14}, which is nearly a decade below the canonical saturation threshold of $\log L_X/L{bol}\sim -3$ \citep[see, e.g.][]{Wright.etal:11}.

LHS~248 was observed using the {\it Chandra} HRC-S on 2018 May 22 for a net exposure of 10~ks.  The target was acquired 
with an offset relative to the on-axis telescope aim point of +5.03 arcmin in order to place the center of the telescope dither pattern on the boundary between the thick and thin Al filter regions. 

An important aspect of the observation is that the filter resides above the detector surface by about 12~mm, and a point source at this height above the focal plane has a size of approximately 30 arcsec. The observation therefore employed a special $128\arcsec \times 16\arcsec$ (full width) dither pattern that was enlarged relative to the standard $40\arcsec \times 40\arcsec$ dither in the spacecraft $Y$ axis, along the long axis of the HRC-S, and compressed in the orthogonal  spacecraft $Z$ axis\footnote{Spacecraft $Y$ and $Z$ axes are aligned with ``chip y'' and ``chip x'', respectively, in HRC-S detector coordinates}.  
This was done so as to minimise the time spent in transit over the thick/thin boundary and to maximise the exposure when the source was entirely either in the thick or thin Al regions.  

\begin{figure}{}
\center
\includegraphics[width=0.46\textwidth]{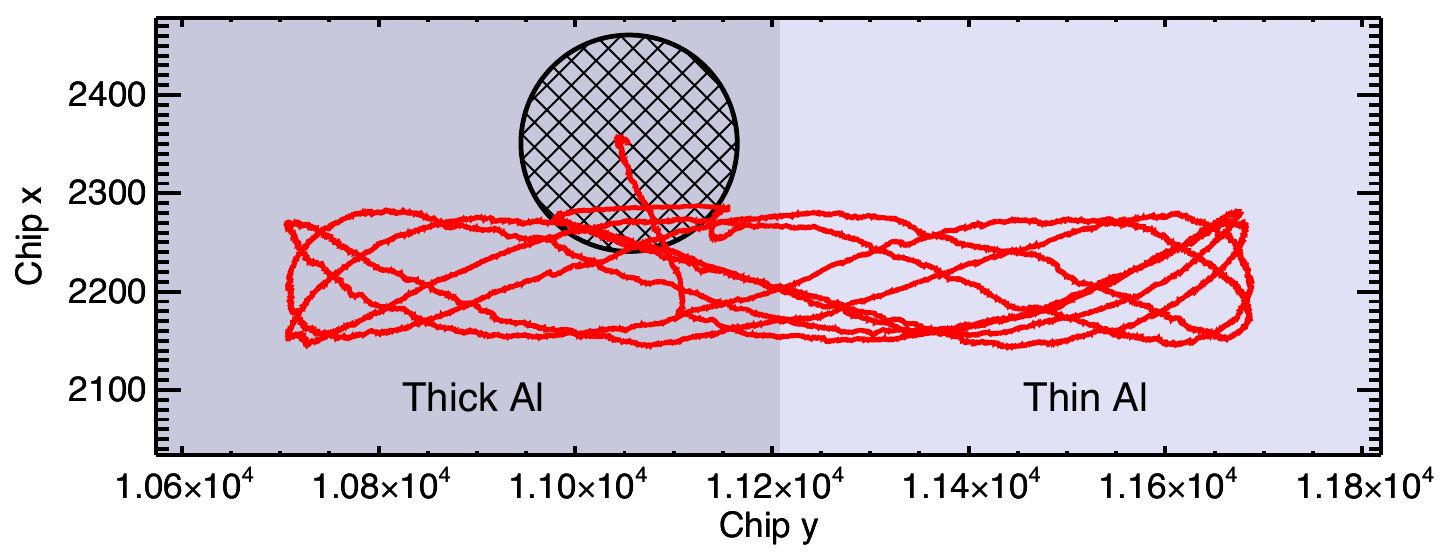} 
\caption{The dither pattern followed by LHS~248 in HRC-S detector coordinates.  The hashed circle represents the size of the converging light cone at the height of the UVIS above the focal plane and is centred on the acquisition position of the target at the beginning of the observation.  Note that the detector {\em Chip y} coordinate is oriented in the opposite direction to the spacecraft coordinates  noted in Figure~\ref{f:hrcs}.
}
\label{f:dither}
\end{figure}

The basis of the observational method employing the spacecraft dither to sample the stellar emission on the different filters is that frequent sampling better accounts for source variability that can occur on the observation timescale than, for example, dividing the exposure time into two continuous pieces. Ideally, the dither period should be as short as possible.  In practice, the shortest dither period achievable is limited by a maximum allowable spacecraft dither speed. In order to accommodate the longer $Y$ axis dither and not violate maximum spacecraft dither rates, the dither period in this axis had to be lengthened from the default 1087s to 2647s.
The dither pattern followed during the observation is illustrated in detector coordinates in Figure~\ref{f:dither}.


{}

\begin{figure}{}
\center
\includegraphics[width=0.46\textwidth]{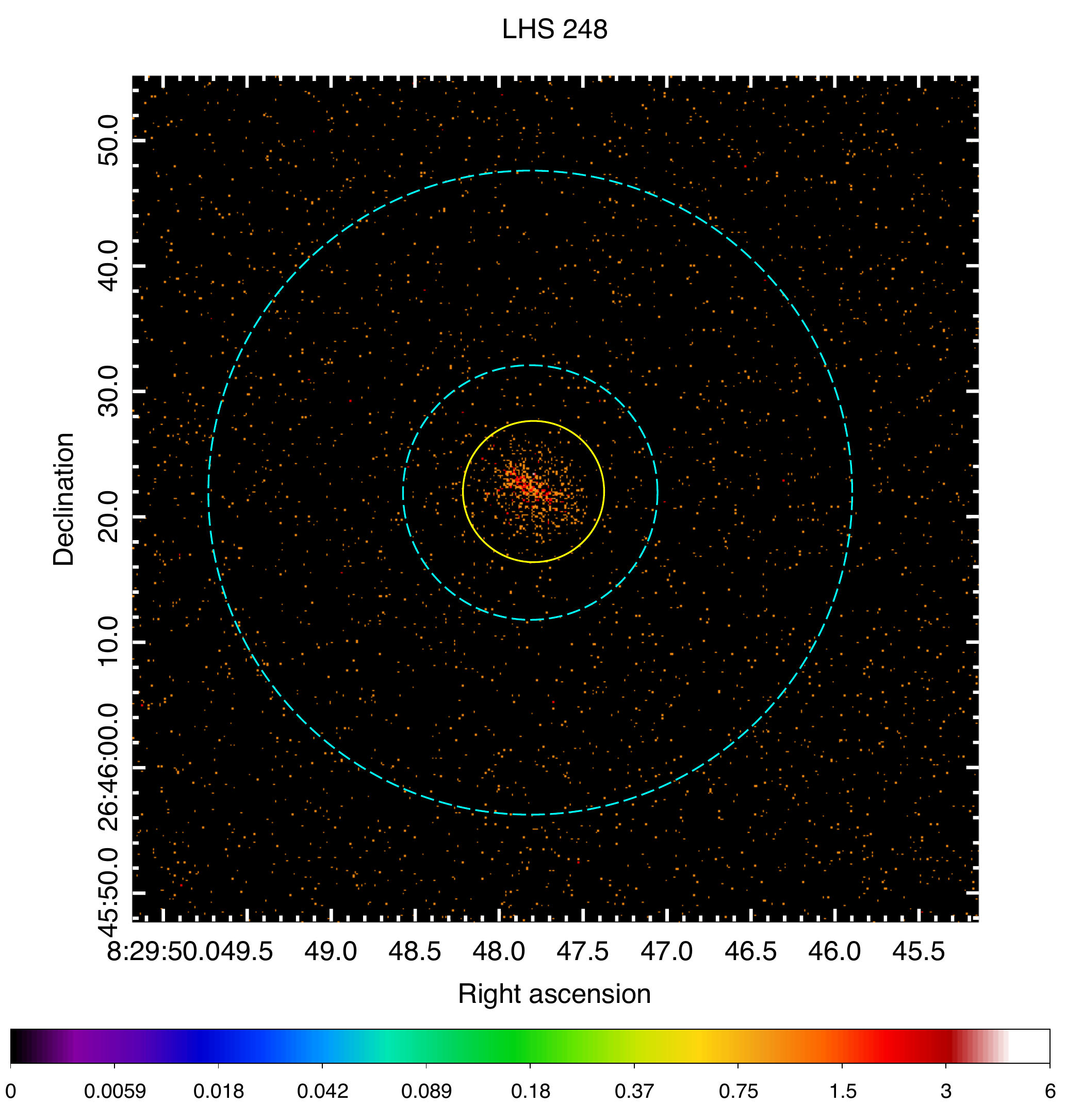} 
\caption{HRC-S image of LHS~248 in sky coordinates illustrating the source and background signal extraction regions employed in the analysis.  The asymmetric nature of the distribution of photon counts is a result of the source being observed off-axis.
}
\label{f:ds9}
\end{figure}

The sky image of the detected events is illustrated in Figure~\ref{f:ds9}, together with the circular and annular regions used to extract the source and background signals, respectively.  The source photons are spread over a distinctly larger detector area than for sources observed on-axis as a result of the degradation of the telescope point spread function with increasing off-axis angle. The source extraction region has a radius of $5.6\arcsec$, with the annular background region having an area 17.6 times larger. 

{}

\subsection{Proxima Centuri}

Proxima is an M5.5 dwarf with a mass of approximately $0.12 M_\odot$ \citep[][and references therein]{Boyajian.etal:12,Delfosse.etal:00} and a rotation period of about 83 days  \citep{Benedict.etal:98,Kiraga.Stepien:07,Savanov:12, Suarez_Mascareno.etal:15}. While of slightly earlier spectral type than LHS~248, its proximity has allowed for detailed study from infrared to X-ray wavelengths and, despite its  relatively slow rate of rotation, its activity level is not too dissimilar from the latter star.  We use it here as a learning example to begin to understand the structure of the coronae of low-mass M dwarfs.

We use high resolution {\it Chandra} observations of Proxima Centauri in the analysis described below in order to provide a simple test of our idealized emission measure distribution approach. Proxima was observed using the High Energy (HETG; \citealt{Canizares.etal:00}) 
and Low Energy (LETG; \citealt{Brinkman.etal:00}) transmission grating spectrometers in 2010 and 2017, respectively (see Table~\ref{t:obs}).  The HETG observation used the Advanced CCD Imaging Spectrometer spectroscopic detector (ACIS-S), while the LETG observations employed the HRC-S.

\begin{table*}[htp]
\caption{Details of {\it Chandra} observations used in this study}
\label{t:obs}
\begin{center}
\begin{tabular}{lccccr}
\hline
Target & ObsID & Instrument &  $t_{\rm start}$ & $t_{\rm stop}$ & Exp.~(s) \\ \hline     \hline
LHS 248 & 20165 & HRC-S & 2018-05-22 07:17:04 & 2018-05-22 10:30:12 & 10065 \\
Proxima & 12360 & HETG+ACIS-S & 2010-12-13 00:19:15 & 2010-12-13 22:56:56 & 78234 \\
Proxima & 19708 & LETG+HRC-S & 2017-05-31 16:25:42 & 2017-06-01 05:28:58 & 44349 \\
Proxima & 20073 & LETG+HRC-S & 2017-05-15 23:29:05 & 2017-05-16 11:22:33 & 40245 \\
Proxima & 20080 & LETG+HRC-S & 2017-05-18 09:21:04 & 2017-05-19 00:28:41 & 51851 \\
Proxima & 20084 & LETG+HRC-S & 2017-06-03 04:45:24 & 2017-06-03 13:51:01 & 29479 \\
\hline
\end{tabular}
\end{center}
\label{default}
\end{table*}%

\section{Analysis}
\label{s:anal}

\subsection{LHS 248 Light Curves}

\begin{figure}{}
\center
\includegraphics[width=0.46\textwidth]{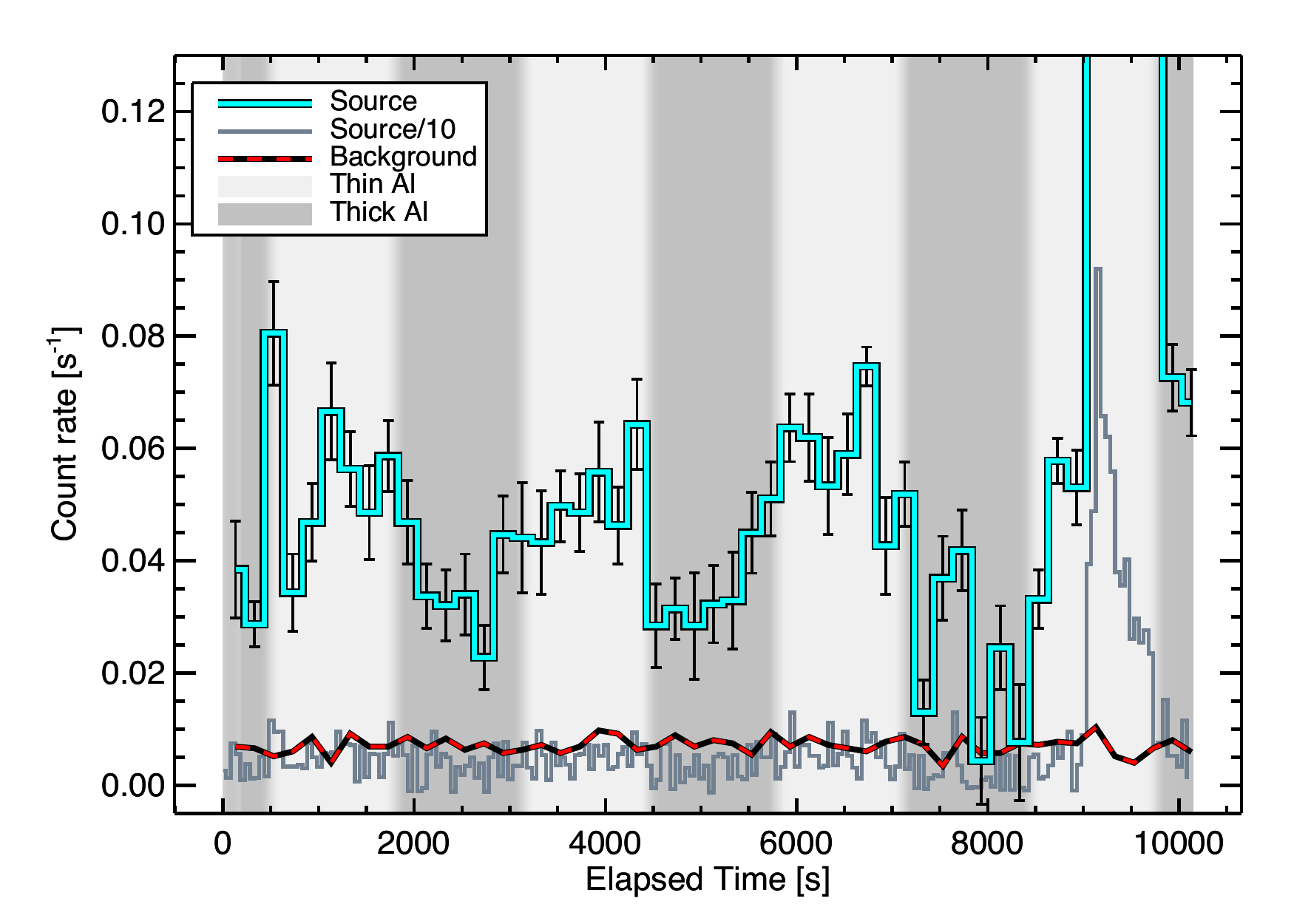} 
\caption{The HRC-S light curve of LHS~248 binned at 200s intervals.  Also shown are the background light curve with the same binning, and the source count rate in 50s bins divided by 10 to illustrate the flare that occurred about 9000s into the observation.  The shaded background represents the times at which the source was on thick or thin Al filter regions. The transitions between these regimes assumed the defocused source could be represented by a uniformly illuminated circle with a radius of 15~arcsec.
}
\label{f:lc}
\end{figure}

The source and background HRC-S light curves of LHS~248 binned at intervals of 200s are illustrated in Figure~\ref{f:lc}, together with the times at which the source was on thick and thin Al filter regions.  This latter calculation assumed that the source was a uniformly illuminated circle and had a radius of $15\arcsec$ at the height of the UVIS above the detector.  While the defocused source is likely to be slightly more donut-shaped than a uniformly illuminated circle, this will have only a second order effect on the light curve shape as the source crosses the thick/thin filter boundary, and does not factor into the selection of intervals during which it lies entirely on either the thick or thin Al region. 

At approximately 9000s from the start of the observation, a large flare was observed in which the count rate was seen to rise by a factor of about 13 over the quiescent value.  This is brought out more clearly in Figure~\ref{f:lc} in the light curve when binned at 50s intervals and divided by a factor of 10.  Since this flare occurred exclusively on the thin Al filter, we excluded it from the analysis described below.

 Figure~\ref{f:lc} demonstrates that, while the source was subject to some stochastic variability, the average count rate when on the thin Al filter was always larger than on the thick Al region, as expected.

{}

\begin{figure}
\center
\includegraphics[width=0.46\textwidth]{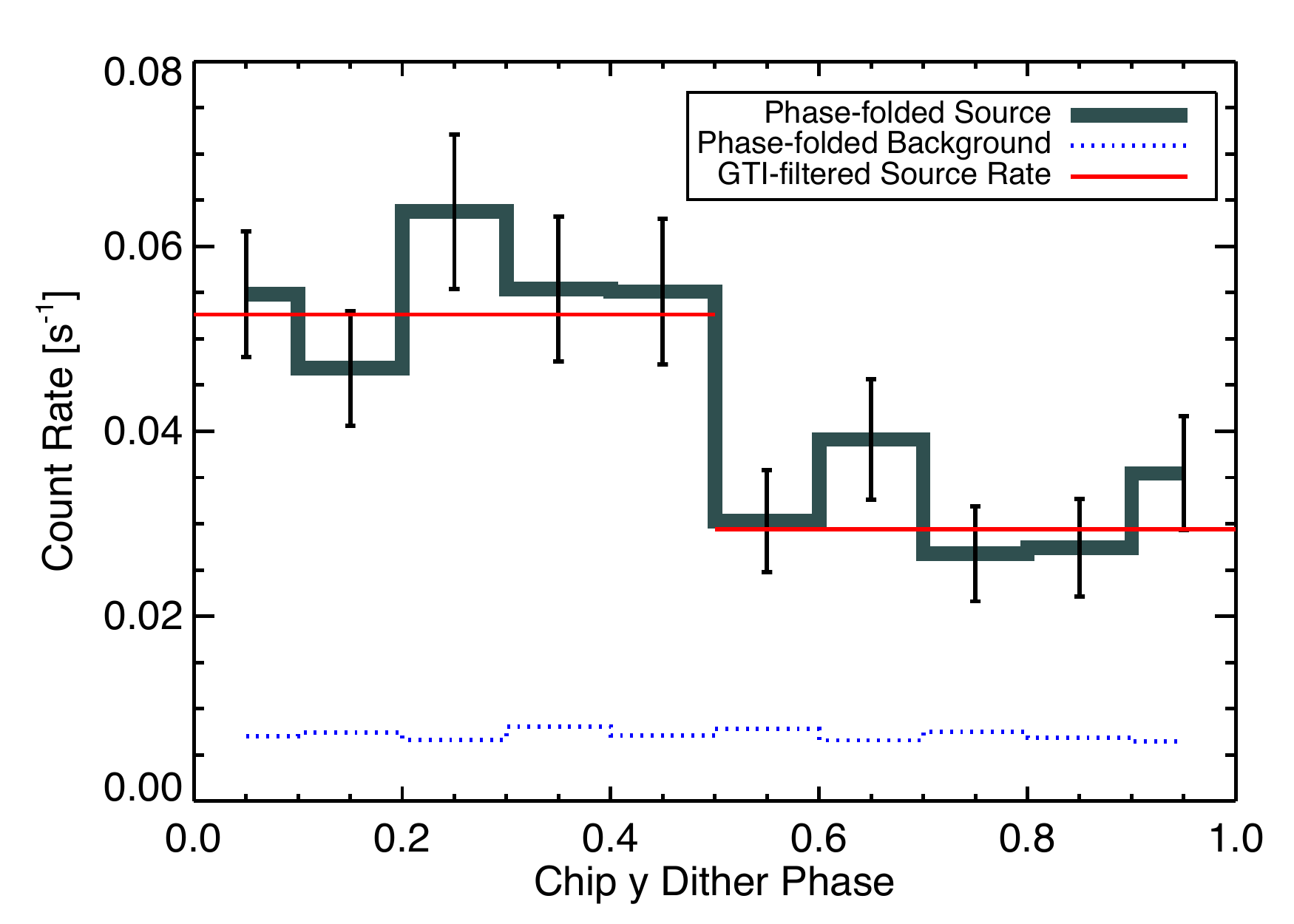} 
\caption{Source and background count rates as a function of the $Chip \; y$ dither phase excluding the large flare toward the end of the observation. The mean count rates for the exclusively thick or thin ``good time intervals'', for which the times during which the source was crossing the thick/thin boundary have been filtered out, are also indicated.
}
\label{f:pfold}
\end{figure}

The source and background light curves are shown as a function of the dither phase in Figure~\ref{f:pfold}.    The average source count rates over the whole observation, excluding the flare and time intervals during which the source was in transition between one filter region and the other---i.e. when source photons were simultaneously incident on both sides of the filter boundary---were $0.0294\pm 0.0031$ (thick), $0.0526\pm0.0040$ (thin), for a count rate ratio thin/thick$\, =1.79\pm0.24$.


\subsection{Assuming an Isothermal Coronal Plasma}
\label{s:isotherm}

The HRC-S detector has essentially no intrinsic energy resolution such that the discrimination between EUV and X-ray photons can only be made by comparing the source count rates in the different filter regions.  We proceed by employing optically-thin plasma radiative loss models in order to, firstly, examine how the thick to thin filter count rate ratio is sensitive to the temperature and abundances in an idealized isothermal corona.  In Sect.~\ref{s:demgrid} below we examine 
the extent to which different shapes of simple parameterized coronal emission measure distributions informed by available observations are consistent with the data.  All the calculations presented here were performed with the IDL\footnote{Interactive Data Language \copyright Harris Geospatial Solutions}-based {\em Package for INTeractive Analysis of Line Emission}\footnote{\sc{PINTofALE} is freely available at http://hea-www.harvard.edu/PINTofALE/ } ({\sc PINTofALE}).

While both solar and stellar coronae are well known to comprise multi-thermal plasmas over a wide range of temperature, it is still useful to examine the extent to which the HRC-S thick and thin Al data are sensitive to isothermal plasma temperature and the assumed chemical abundance mixture.  To this end, synthetic spectra were computed within PINTofALE for isothermal temperatures from $10^5K$ to $10^8K$ using emissivities from the {\it CHIANTI} database version 7.1.5. We computed two sets of spectra, one for the solar abundances of \citet{Grevesse.Sauval:98}, and the other for an ``inverse first ionization potential'' (inverse FIP) chemical composition.  
The latter reflects the now extensive observational evidence that in active stars the abundance of elements with low first ionization potential (FIP; $< 10$eV) appear to be depleted relative to elements with high FIP by factors of two or more.  This is in contrast to the picture of lower activity stars like the Sun, in which low FIP elements appear to be enhanced \citep[see, for example,][]{Drake.etal:97,Drake:03,Laming:15,Wood.etal:18}.  These coronal abundance patterns also appear to depend on spectral type, with later K and M stars exhibiting a more inverse FIP pattern and earlier types tending towards a solar-like FIP effect \citep{Wood.etal:18}.  

An {\it XMM-Newton} study of four active mid-M dwarfs by \citet{Robrade.Schmitt:05} found all to be characterised by an inverse FIP abundance pattern in which low FIP elements such as Mg, Si, and Fe were depleted by about a factor of 2 relative to solar photospheric values while high FIP elements, such as Ne, C, O, and N were relatively enhanced by up to a factor of 2.  \citet{Robrade.Schmitt:05} noted that the low FIP deficiency might simply reflect the stellar photospheric metallicity, leaving the high FIP elements enhanced.  For our inverse FIP abundance set we adopted a simple pattern in which low FIP element abundances were reduced by a factor of 2 relative to the \citet{Grevesse.Sauval:98} mixture.

Synthetic spectra were convolved with the instrument effective area curves to obtain predicted count rates.  We first investigated the effect of interstellar absorption for application to sources in general at arbitrary distances by computing the filter count rate ratios for a range of interstellar medium absorbing columns ranging from $10^{18}$ to $10^{22}$~cm$^{-2}$; ratios for lower column densities are essentially identical to that for $10^{18}$~cm$^{-2}$.  
The absorbing column toward LHS~248 is not known with any great degree of precision.  According to the local interstellar cloud model of \citet{Redfield.Linsky:00}\footnote{http://lism.wesleyan.edu/cgi-bin/distlic.cgi}, the distance to the edge of the cloud in the line of sight toward LHS~248 is 2.3~pc. They estimate an average neutral hydrogen density of 0.1 atoms cm$^{-3}$, for a column density of  $7.1\times 10^{17}$~cm$^{-2}$.  Beyond the local cloud in the same sightline lies the Gemini cloud\footnote{http://lism.wesleyan.edu/cgi-bin/dynlism.cgi} (assuming the structure of the local ISM region comprises discrete clouds; see, e.g., \citealt{Redfield.Linsky:15}), although there is no strong constraint on its distance.  \citet{Redfield.Linsky:08} list a distance of 6.7~pc to the closest star with measured absorption signatures from this cloud.  It is therefore possible that there is no further significant absorption toward LHS~248 above that of the local cloud.  A range of $5\times 10^{17}$---$3\times 10^{18}$~cm$^{-2}$ should then comfortably bracket the actual neutral hydrogen column toward LHS~248.

\begin{figure}{}
\center
\includegraphics[width=0.46\textwidth]{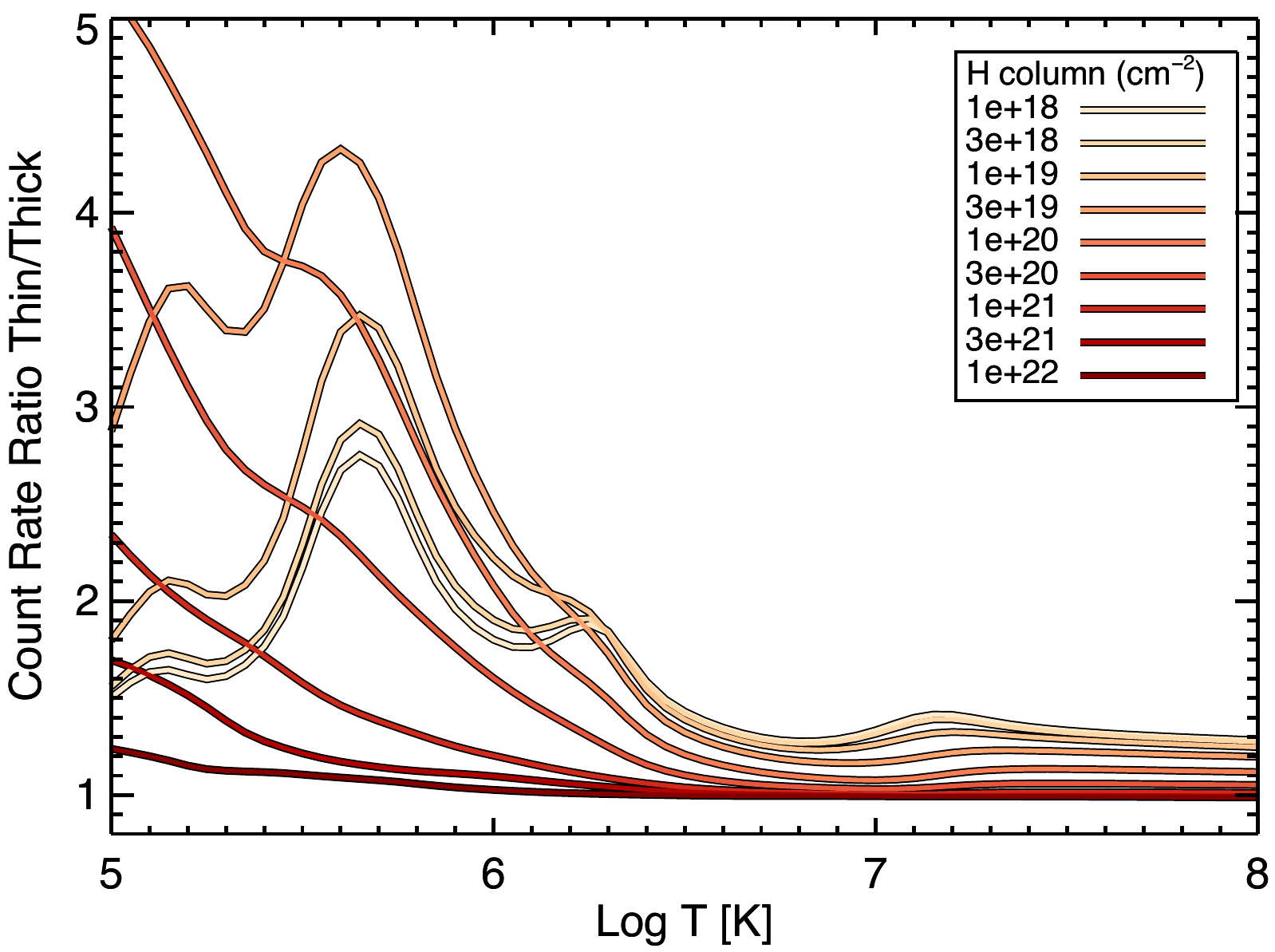} 
\includegraphics[width=0.46\textwidth]{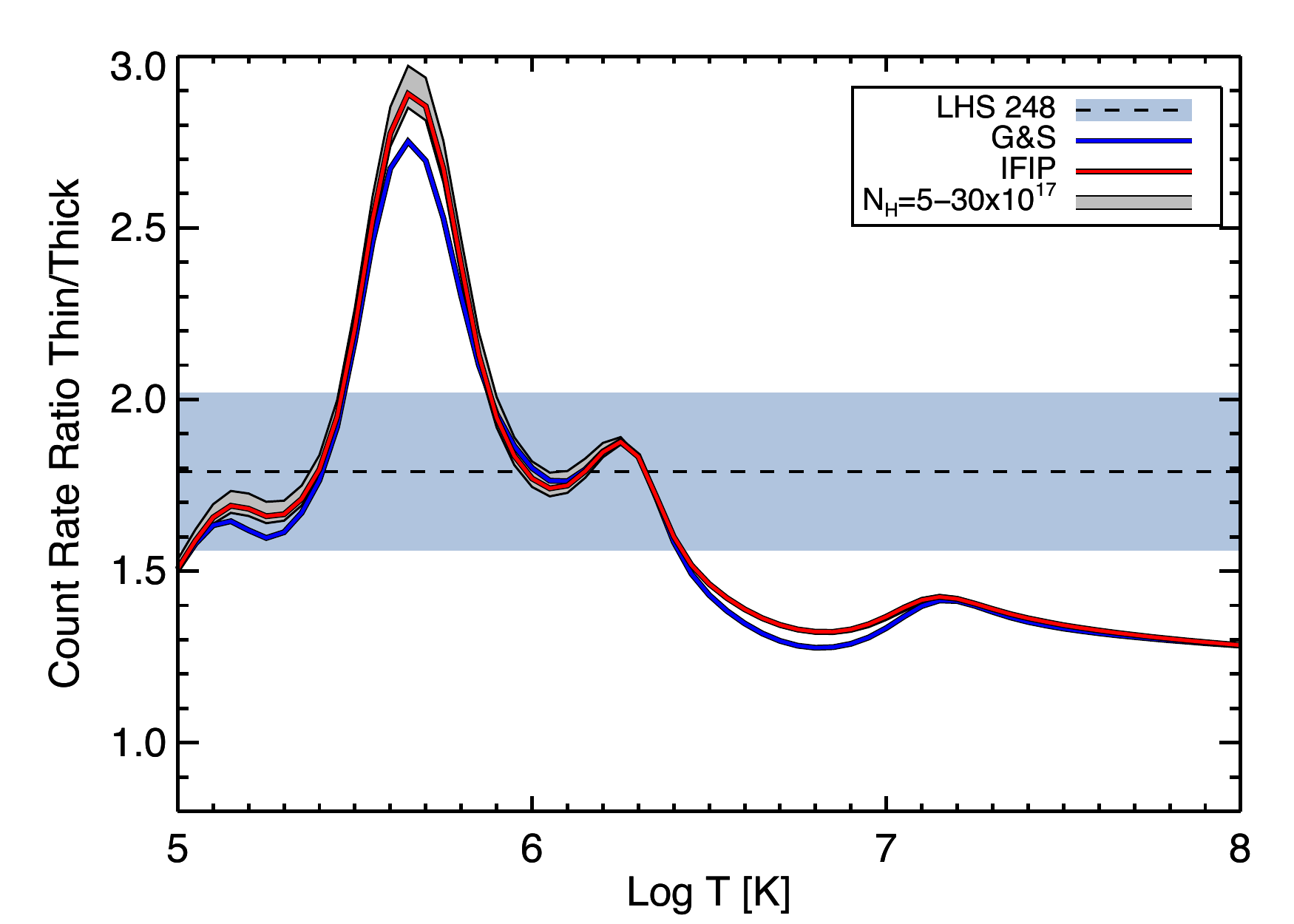} 
\caption{The predicted thin/thick filter count rate ratio for {\em iosothermal} optically-thin plasma radiative loss models. Top: for the solar \citep{Grevesse.Sauval:98} abundance mixture and different values of interstellar medium neutral hydrogen column density.  Bottom:  The count rate ratio for the column density range appropriate to LHS~248, and including both solar (``G\&S'') and inverse FIP (see text) abundance mixtures; the observed ratio for LHS~248 is also indicated.
}
\label{f:filtrat}
\end{figure}

The predicted count rate ratios for the filters as a function of isothermal plasma temperature for the different ISM absorbing columns, and for the solar and inverse FIP compositions for absorption appropriate to LHS~248 are illustrated in Figure~\ref{f:filtrat}. 

{}

\subsection{Coronal Emission Measure Distributions of M dwarfs}

\subsubsection{General Considerations}
\label{s:gen_dem}

We base the analysis of LHS~248 below on simple parameterizations of the coronal emission measure distribution.  In its most simple form, the emission measure as a function of temperature is usually expressed in the form of the differential emission measure (DEM) as a function of temperature, $T$,
\begin{equation}
DEM(T)=n_e^2(T)\frac{dV(T)}{d \log T}.
\label{e:dem}
\end{equation}

Since the advent of high resolution EUV and X-ray spectroscopy of, first, the solar corona corona and, subsequently, stellar coronae, many studies have examined the form of the plasma DEM.  The reader is referred to the following work as a starting point for deeper exploration of the extant literature: \citet{Pottasch:63,
Withbroe:75, Jordan:76,Craig.Brown:76,Bruner.McWhirter:88,Kashyap.Drake:98,Jordan:00} for solar work;  \citet{Drake.etal:95,Sanz-Forcada.etal:02,Sanz-Forcada.etal:03,Huenemoerder.etal:03,Telleschi.etal:05,Wood.etal:18} for example stellar results.

The body of existing work demonstrates some universal aspects of the coronal DEM: from chromospheric temperatures of a few $10^4$~K the DEM decreases by approximately an order of magnitude to a minimum at temperatures of approximately 1--$4\times 10^5$~K; the DEM then rises by a approximately order of magnitude, or more, to a maximum at temperatures between $10^6$--$10^7$~K, beyond which it extends to higher temperatures over either a plateau or a shallow downward slope, before a precipitate decline by several orders of magnitude.  It is possible there is further fine structure in the shape of the DEM in some cases, although assessing the veracity of such structure is far from trivial owing to the nature of the ill-constrained integral inversion problem of inferring the DEM from observed spectra  \citep[see, e.g.,][]{Craig.Brown:76,Kashyap.Drake:98}.

\subsubsection{The case of Proxima: {\it Chandra} HETG Spectrum}
\label{s:proxhetg}

While existing stellar DEM data exhibit common characteristics, there is essentially no detailed spectral information for late M dwarfs with which to verify that such characteristics do extend to the lowest mass stars.  Here, we examine the {\it Chandra} HETG spectrum of Proxima Centuri in order to help establish the form of emission measure distribution to adopt for our target LHS~248. 
The analysis employed a parameter estimation approach in which the observed spectrum was fitted to a multi-thermal, optically-thin plasma radiative loss model whose emission measure distribution at each temperature was allowed to vary.

\begin{figure}
\center
\includegraphics[width=0.46\textwidth]{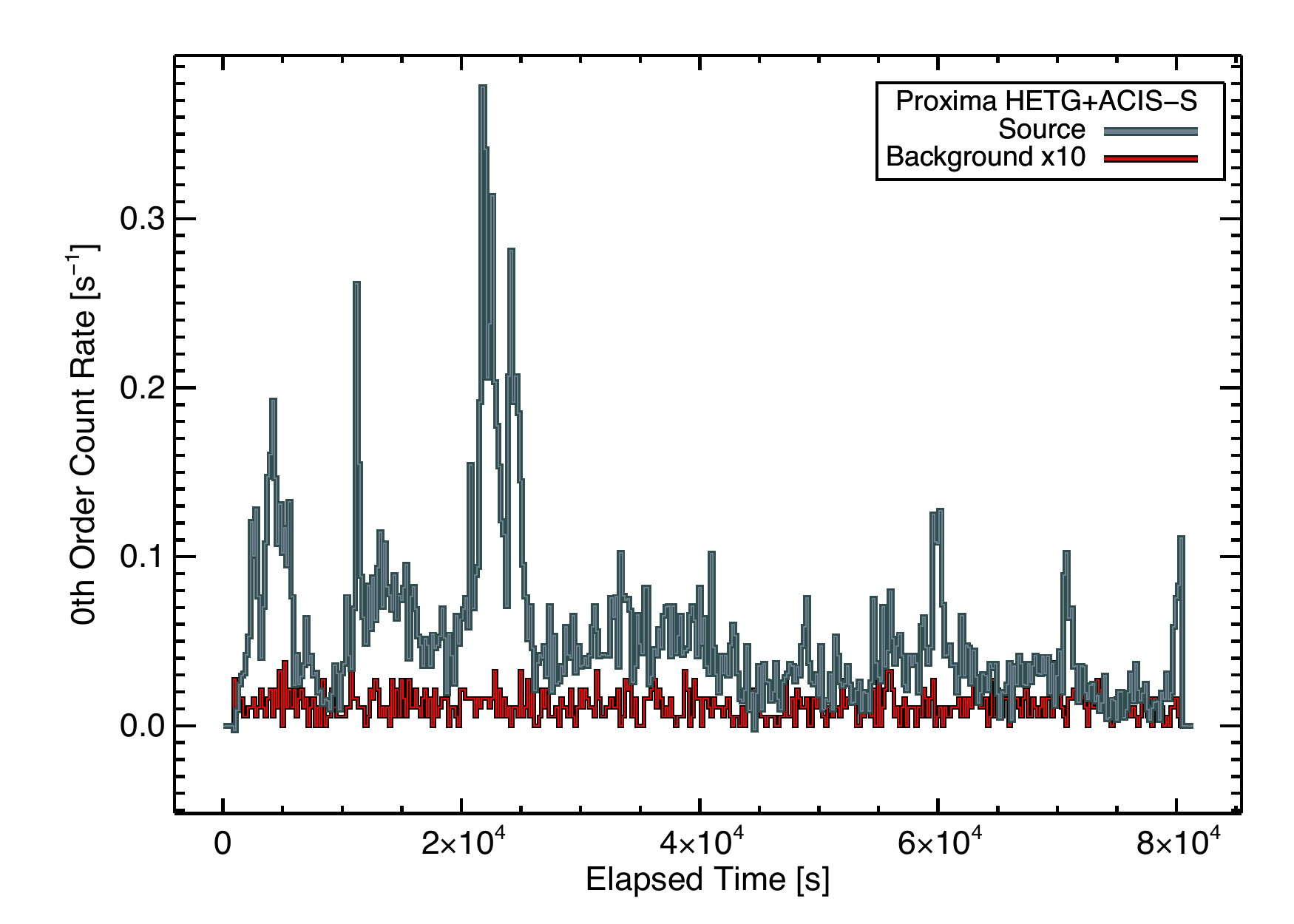} 
\includegraphics[width=0.46\textwidth]{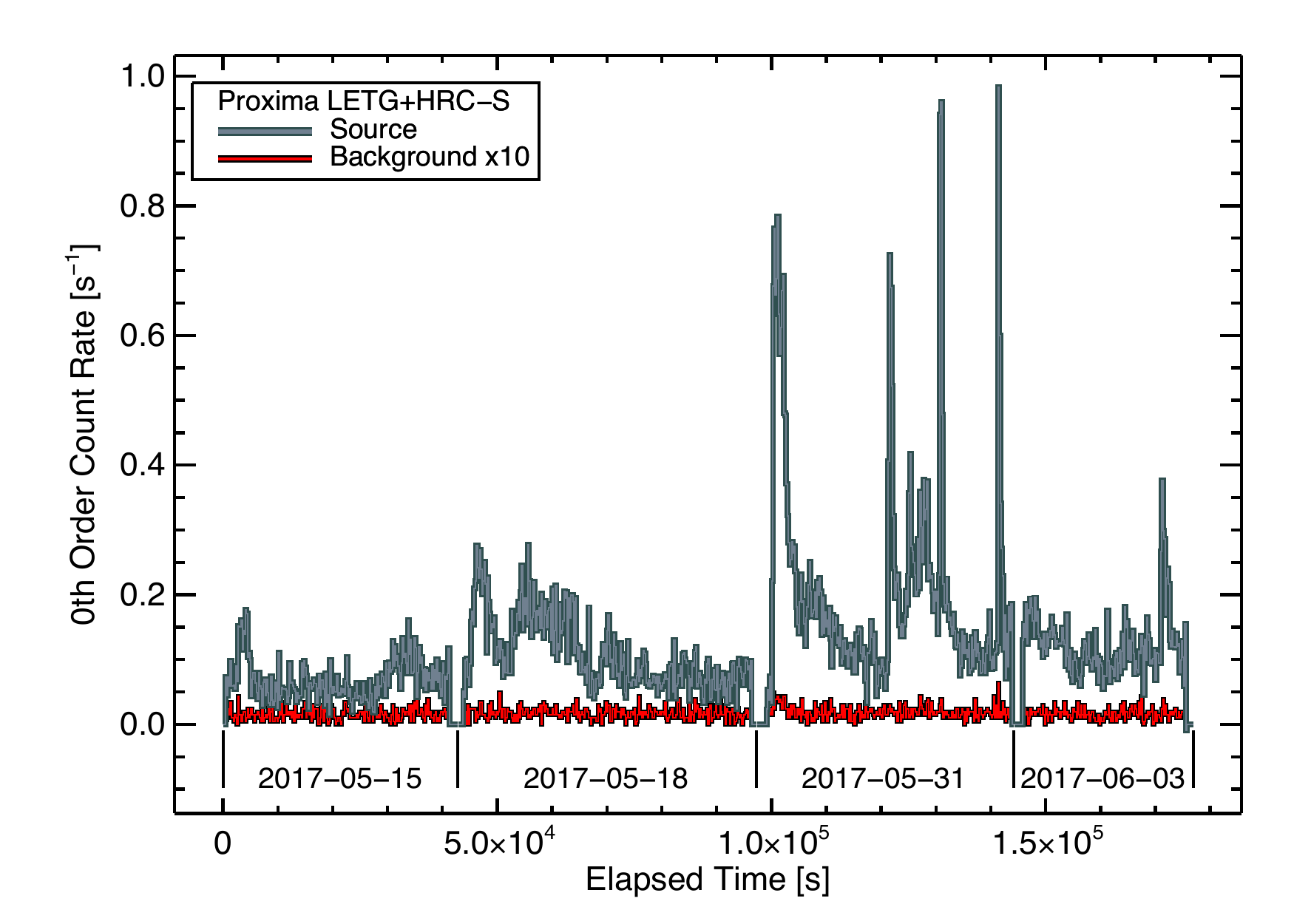} 
\caption{X-ray light curves of Proxima observed with the {\it Chandra} HETG+ACIS-S (top), and LETG+HRC-S (bottom). Light curves were obtained from the 0th orders in both cases, and were binned at intervals of 200s. The HETG+ACIS-S 0th order light curve has been corrected for pile-up (see text).
}
\label{f:lc_prox}
\end{figure}

Proxima is the most well-known flare star in the sky, and in order to understand the context of the extracted spectra upon which our analysis is based we first extracted the dispersed photon events in 0th order to compute the X-ray light curve and assess flaring activity.  An approximate correction for the effects of ``pile-up''---when photon events are lost or incorrectly counted due to multiple photon interactions at the same detector location during a single CCD frame---using pile-up fraction estimates from the {\it Portable, Interactive Multi-Mission Simulator}\footnote{http://cxc.harvard.edu/toolkit/pimms.jsp} was made to the extracted light curve.  The resulting corrected light curve is illustrated in Figure~\ref{f:lc_prox}. Pile-up fractions varied from a maximum of approximately 30\%\ for the very peak of the flare at about 20~ks into the exposure to a minimum of about 2\%\ for the lowest count rates and a more typical 6\%\ for count rates of $\sim 0.05$~count~s$^{-1}$.  The observation is characterised by a moderate, and likely typical, amount of flaring \citep[see, eg, the survey of Proxima X-ray observations by][]{Wargelin.etal:17}, in which several events are observed to reach count rates an order of magnitude above the apparent quiescent rate.

Parameter estimation was undertaken using the {\it Sherpa} fitting engine within CIAO.  Multiple Astrophysical Plasma Emission Code (APEC\footnote{http://www.atomdb.org/}) thermal plasma 
models were adopted to represent the DEM on a fixed temperature grid, each with a common fixed abundance pattern whose absolute normalization---the metallicity---was free to vary.  The temperature grid spanned the range $\log T=6.25$--7.5 in intervals of 0.25, for a total of 6 thermal plasma components.

While the abundances could in principle be fitted simultaneously with the temperature structure, we found that the significantly larger number of free parameters required posed problems for the stability of the solution.  The solar abundances of  \citet{Grevesse.Sauval:98}, modified to approximate the ``inverse FIP'' pattern by increasing by a factor of two the abundances of C, N, O, Ne and Ar, as described above in Sect.~\ref{s:isotherm}, were adopted.

\begin{figure*}{}
\center
\includegraphics[width=0.54\textwidth]{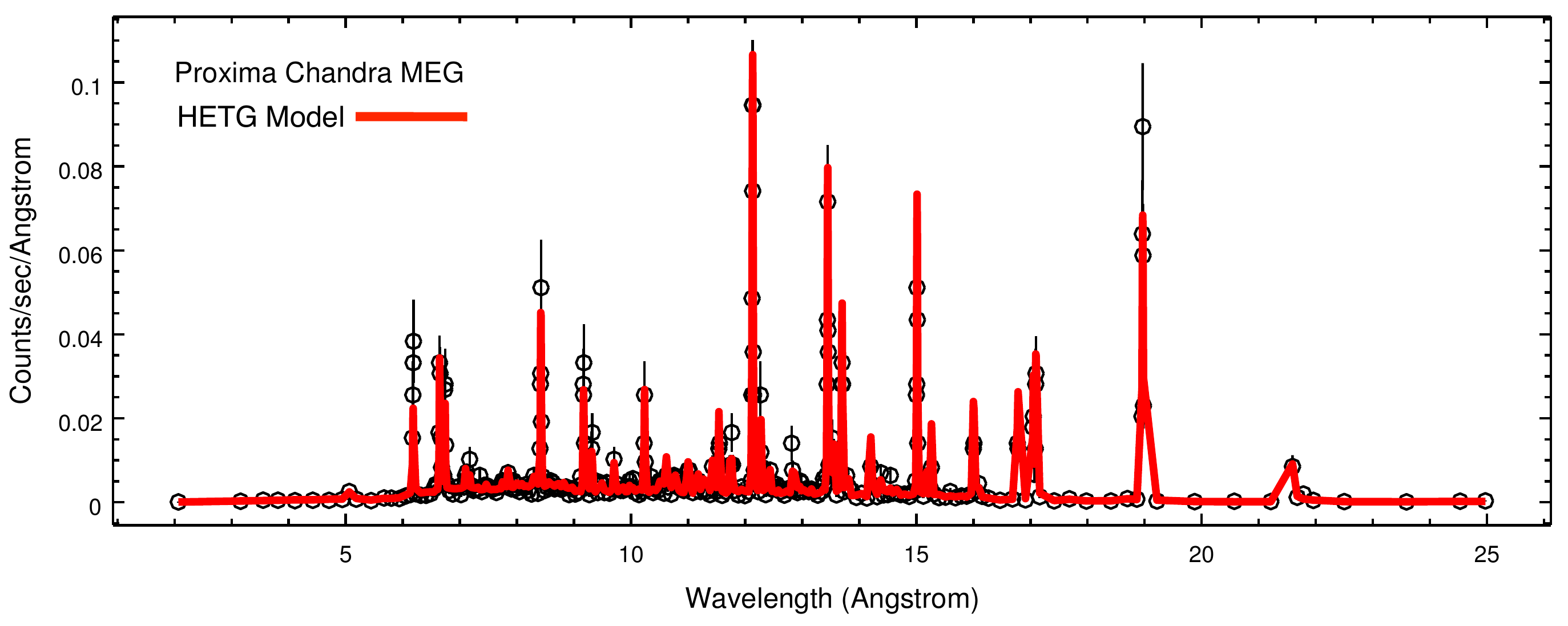} 
\includegraphics[width=0.54\textwidth]{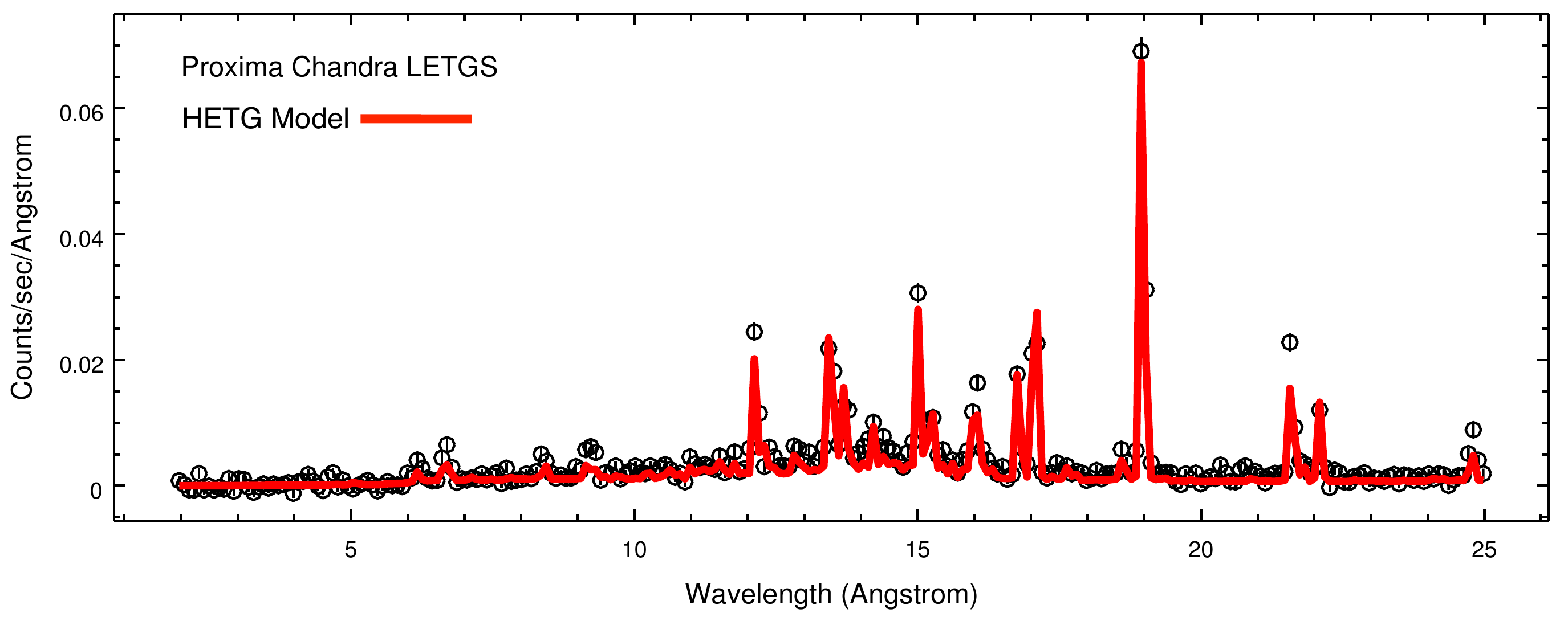} 
\includegraphics[width=0.54\textwidth]{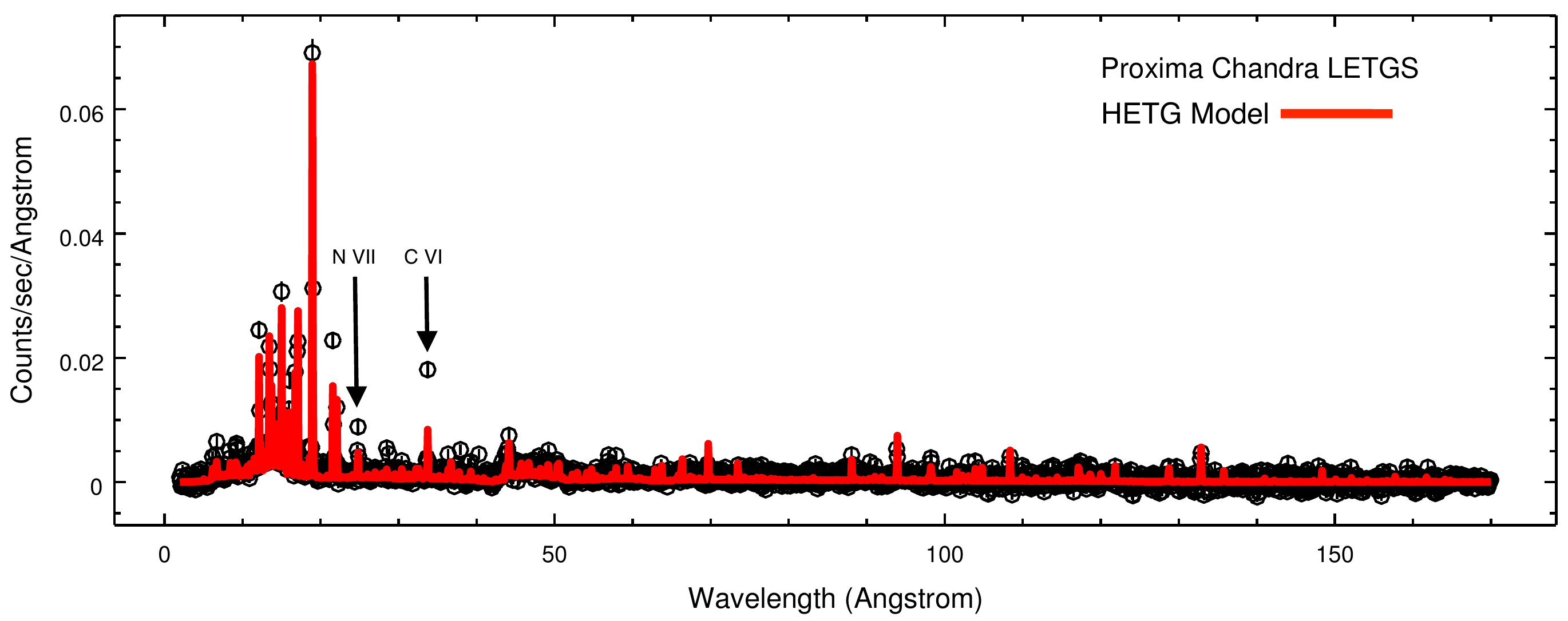} 
\includegraphics[width=0.54\textwidth]{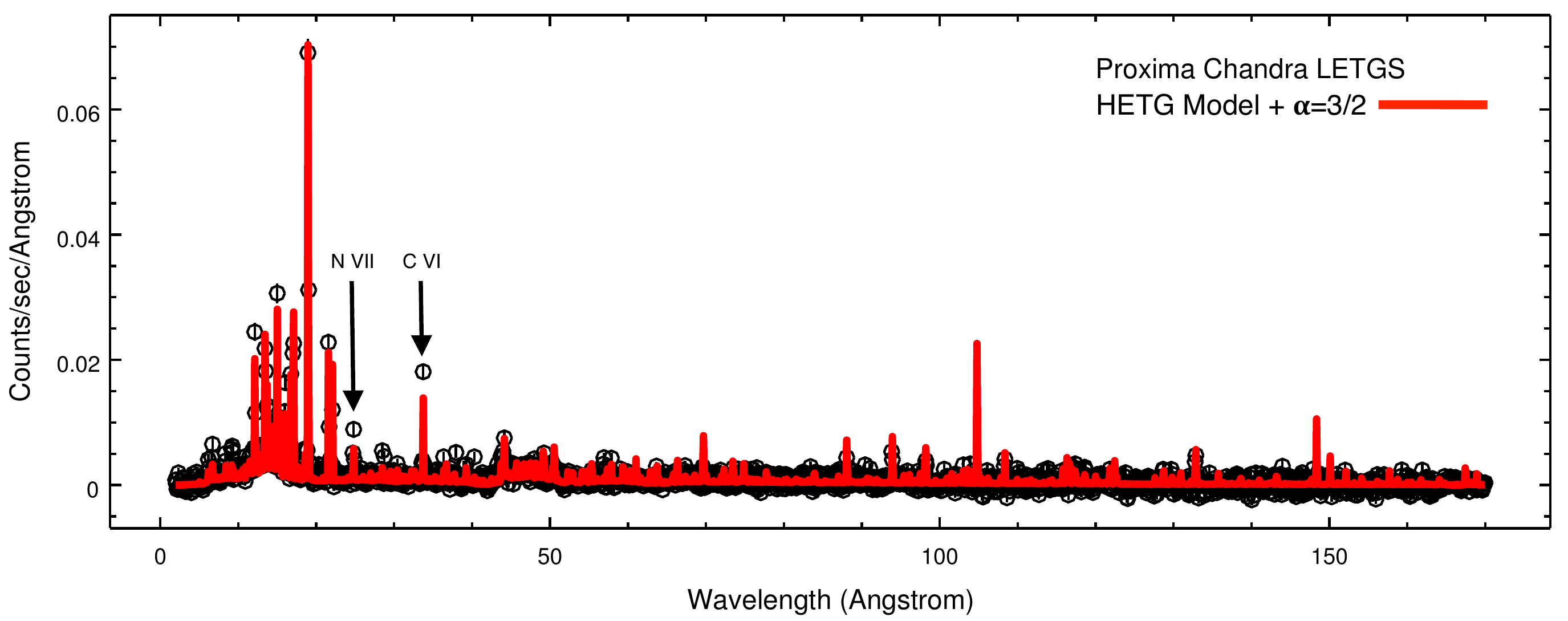}
\includegraphics[width=0.54\textwidth]{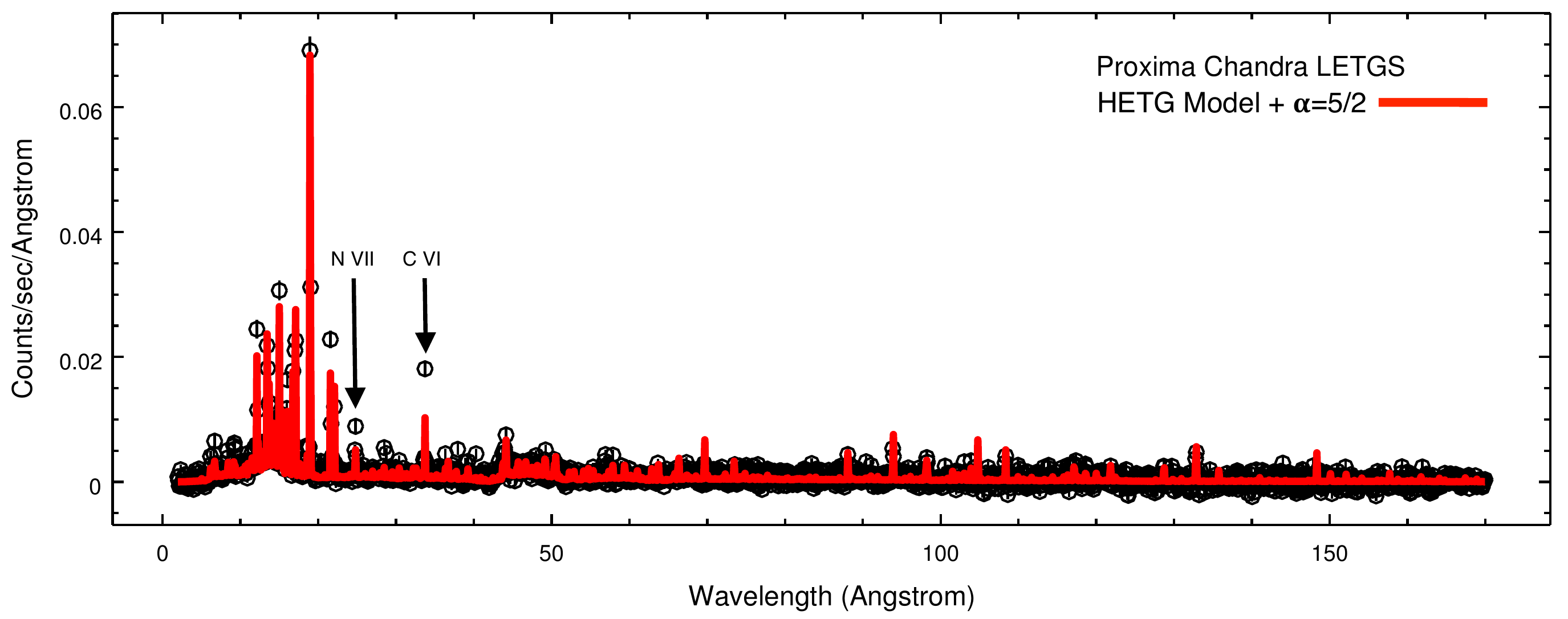} 
\caption{The {\it Chandra} HETG (top panel) and LETG (lower four panels) spectra of Proxima overlaid with the best-fit HETG model spectrum (see text for details).  The lower three panels illustrate the full range of the LETG+HRC-S spectrum and show the effect of changing the cool ``tail" of the DEM.  The first of these (middle panel) illustrates the HETG model with no additional cool plasma components.  The lower two show the HETG model augmented by cool DEM extensions that follow the power law relation $DEM(T) \propto T^\alpha$ with $\alpha=3/2$ (fourth panel down) and $\alpha=5/2$ (bottom panel).  In order to aid the visualizations of the model comparisons, spectra have been rebinned. 
}
\label{f:proxfits}
\end{figure*}{}

Since the high-resolution {\it Chandra} spectra of Proxima comprise many bins with few counts, the Cash statistic ($\mathcal C_r$; \citealt{Cash:79}) was employed for minimization of model deviations from the data.  This statistic is valid in the Poisson regime of low numbers of counts in which $\chi^2$ approaches, relying on Gaussian uncertainties, are inapplicable.   This allows the data to be analyzed without further grouping of neighboring bins, and at full spectral resolution.  

The HETG High Energy Grating (HEG) spectrum was fitted over the wavelength range 1.5--16~\AA, while fits to the Medium Energy Grating (MEG) spectrum were restricted to 2--25~\AA. Both HEG and MEG spectra were initially fit simultaneously, but the low signal-to-noise ratio of the HEG spectrum lead to difficulties in obtaining convergence and final adopted results were based on the best-fit to the MEG spectrum. The best-fit model yielded a reduced statistic of 0.6 and a formal metallicity of $[M/H]=-0.37\pm 0.04$, expressed in the conventional logarithmic bracket notation. 

\begin{figure}
\center
\includegraphics[width=0.46\textwidth]{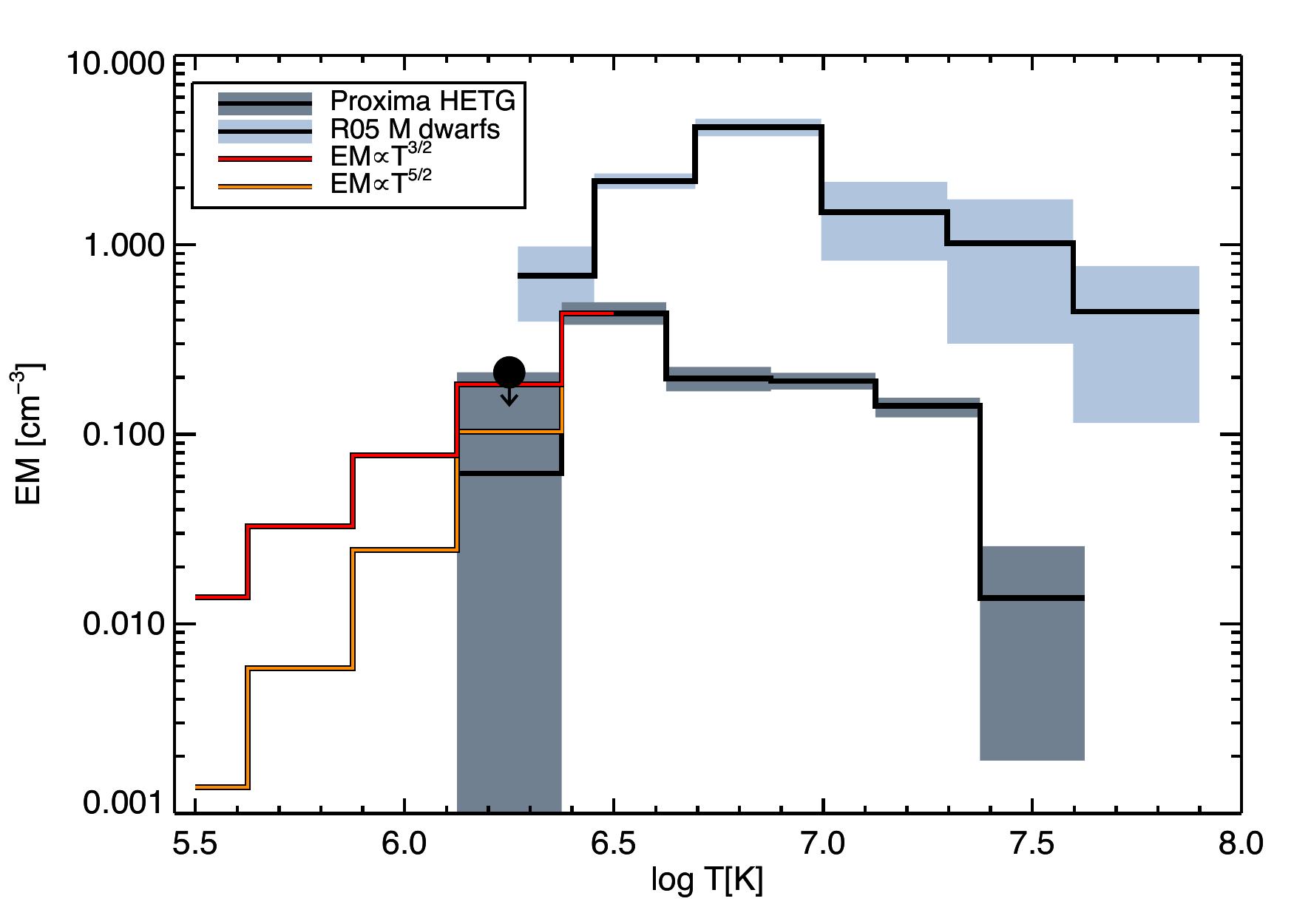} 
\caption{Emission measure distributions derived from HETG+ACIS-S observations of Proxima, together with the mean DEM from the {\it XMM-Newton} M dwarf study of \citet[][R05]{Robrade.Schmitt:05}.  The grey shaded regions represent uncertainties in the form of 90\%\ confidence intervals. The cool extensions of the Proxima DEM tested on the LETG+HRC-S spectrum of Proxima are also illustrated.
}
\label{f:dem}
\end{figure}

The best-fit model is illustrated superimposed on the MEG spectrum in top panel of Figure~\ref{f:proxfits}, while the resulting emission measure distribution is shown in Figure~\ref{f:dem}.  

\subsubsection{Proxima Cen LETGS spectrum}
\label{s:proxletg}

A global model fit of the {\it Chandra} LETGS spectrum similar to that performed for the HETG unfortunately proved infeasible due to the low signal-to-noise ratio in the vast majority of the spectral bins. However, it was instead immediately instructive to simply superimpose the spectrum computed for the HETG best-fit parameters.  This overlay is illustrated in Figure~\ref{f:proxfits}. Note that no scaling or adjustment to any parameters has been made, except to scale back the C and N abundances to their solar ratios, [C/Fe]=0 and [N/Fe]=0 for illustrative reasons explained below.  The agreement at wavelengths $< 25$~\AA\ is quite remarkable, indicating that the average activity level of the two observations, flares included, was essentially the same.

The coolest temperature constrained by the HETG data was an upper limit to the DEM at logarithmic temperature $\log T=6.25$. 
The longer wavelengths of the LETGS spectral range exhibit lines formed at cooler plasma temperatures than sampled in the HETG range and we can in principle use these lines to better constrain the lower temperature DEM. A glance at the centre panel of Figure~\ref{f:proxfits} reveals a slew of emission lines longward of 50~\AA\ in the model spectrum that appear in generally good agreement with the observed spectrum in the sense that they at least do not exceed the observed flux. 

We next illustrate two similar spectral models computed with cool extensions to the HETG DEM that are illustrated in Figure~\ref{f:dem}.
The form of the DEM extension is fairly arbitrary but based on the generally observed trends noted in Sect.~\ref{s:gen_dem}; we simply extrapolate from the trend in lower-T behavior seen here and found in M dwarfs by \citet[][see Sect.~\ref{s:demgrid} and Figure~\ref{f:dem}]{Robrade.Schmitt:05}, using 
$DEM \propto T^\alpha$, with $\alpha=3/2$ and 5/2. The spectrum corresponding to a shallower power law slope, $\alpha=3/2$, egregiously over-predicts several lines, most notably the O~VI doublets at 104.8~\AA\ and 150~\AA, and Ni~X and XI at 148.6 and 148.4~\AA, respectively.  Instead, the spectrum corresponding to the steeper DEM slope, $\alpha=5/2$, is in quite good agreement with the lines in the longer wavelengths of the LETG spectrum. 

Returning to the elements C and N, whose abundances were not scaled to the inverse FIP pattern adopted for the HETG spectrum fitting for the spectra illustrated in Figure~\ref{f:proxfits}, we note that both their H-like doublets at 33.7~\AA\ and 24.8~\AA\ denoted in the lower three panels are under-predicted by a factor of 2 in the $\alpha=5.2$ spectrum.  The He-like resonance lines at 28.8 and 40.3~\AA\ are similarly under-predicted.  This remains the case even for the elevated cool DEM $\alpha=3.2$ spectrum, although to a slightly lesser extent.
These lines are formed at 1--$2\times 10^6$~K and their under-prediction indicates that the inverse FIP abundance pattern extends to cooler temperatures than represented by the bulk of the HETG spectrum. 

\subsection{A grid of model DEMs for LHS~248}
\label{s:demgrid}

The general emission measure distribution shape that matches {\it Chandra} HETG and LETG spectra of Proxima reasonably well comprises a fairly steep power law slope from $\log T=5.5$ until a peak at $\log T=6.5$, a shallow power law-like decline to $\log T=7.25$, followed by a steep drop off toward higher temperatures. By way of an interesting comparison to Proxima, Figure~\ref{f:dem} also illustrates the {\em average} emission measure distribution for the four active mid-M dwarfs (AD Leo, EV~Lac, AT Mic and EQ Peg) whose {\it XMM-Newton} spectra were analysed by \citet{Robrade.Schmitt:05}. The average of these M dwarf DEMs can be well-approximated, at least within the temperature range probed by the {\it XMM-Newton} data, by two power law components: a cool component rising toward a the temperature of peak emission measure and a declining power law toward higher temperatures. 

Inspired by these similar-looking general DEM characteristics, and lacking detailed spectral information, we proceed to analyse the LHS~248 HRC-S photometry using a grid of model DEMs with a shape guided by the observed DEM characteristics. The general form adopted is 
\begin{align}
\Phi(T)& = \Phi(10^4K)\left(\frac{T}{10^4\,K}\right)^{\alpha_T};  &\; 10^4 K \leq T \leq T_{min} 
\nonumber \\
 & = \Phi(T_{min})\left(\frac{T}{T_{min}}\right)^{\alpha_{LC}}; &\; T_{min} \leq T \leq T_{peak} 
\nonumber \\
 & = \Phi(T_{peak})\left(\frac{T}{T_{peak}}\right)^{\alpha_{UC}}; &\; T_{peak} \leq T \leq T_{sh} \nonumber \\
 & = \Phi(T_{sh})\left(\frac{T}{T_{sh}}\right)^{\alpha_{HT}}; &\; T_{sh} \leq T \leq 10^8 K
\end{align}
where $T_{min}$ is the temperature at which the DEM reaches a minimum, $T_{peak}$ is the temperature at which the DEM peaks, $T_{sh}$ is the maximum temperature reached by the DEM ``plateau'' at its shoulder region, and $\alpha_T$, $\alpha_{LC}$, $\alpha_{UC}$ and $\alpha_{HT}$ are the power law indices for the transition region, the ``lower corona'', the ``upper corona'' and the high temperature tail, describing the slopes to the DEM in the different temperature regions. The value of $\alpha_{LC}$ is always positive, $\alpha_T$ and $\alpha_{HT}$ are always negative, and $\alpha_{UC}$ could in principle be either positive or negative but in our model DEMs takes on a fixed negative value.

In order to limit the number of parameters in the analysis to a manageable set, we use the observed DEMs in Figure~\ref{f:dem} and appearing in the literature (Sect.~\ref{s:gen_dem}) to constrain the general shape and parameter range to investigate.  The temperature of the DEM minimum was assumed to be $\log T_{min}=5.5$, below which a fixed power law was assumed for all DEM models.  Plasma at temperatures below $\log T=5.5$ does not contribute significant emission in our HRC-S bandpasses in the presence of hotter coronal emission and the details of this cooler part of the DEM are not important for our analysis.  The full set of fixed constraints are 
\begin{align}
\log T_{min} &= 5.5 \nonumber \\
\log T_{sh}-\log T_{peak}&=0.75 \nonumber \\
\alpha_T & =-1.33\nonumber \\
\alpha_{UC} & =-0.67 \nonumber \\
\alpha_{HT} &=-10.
\label{e:dem_fixed}
\end{align}

\begin{figure}{}
\center
\includegraphics[width=0.46\textwidth]{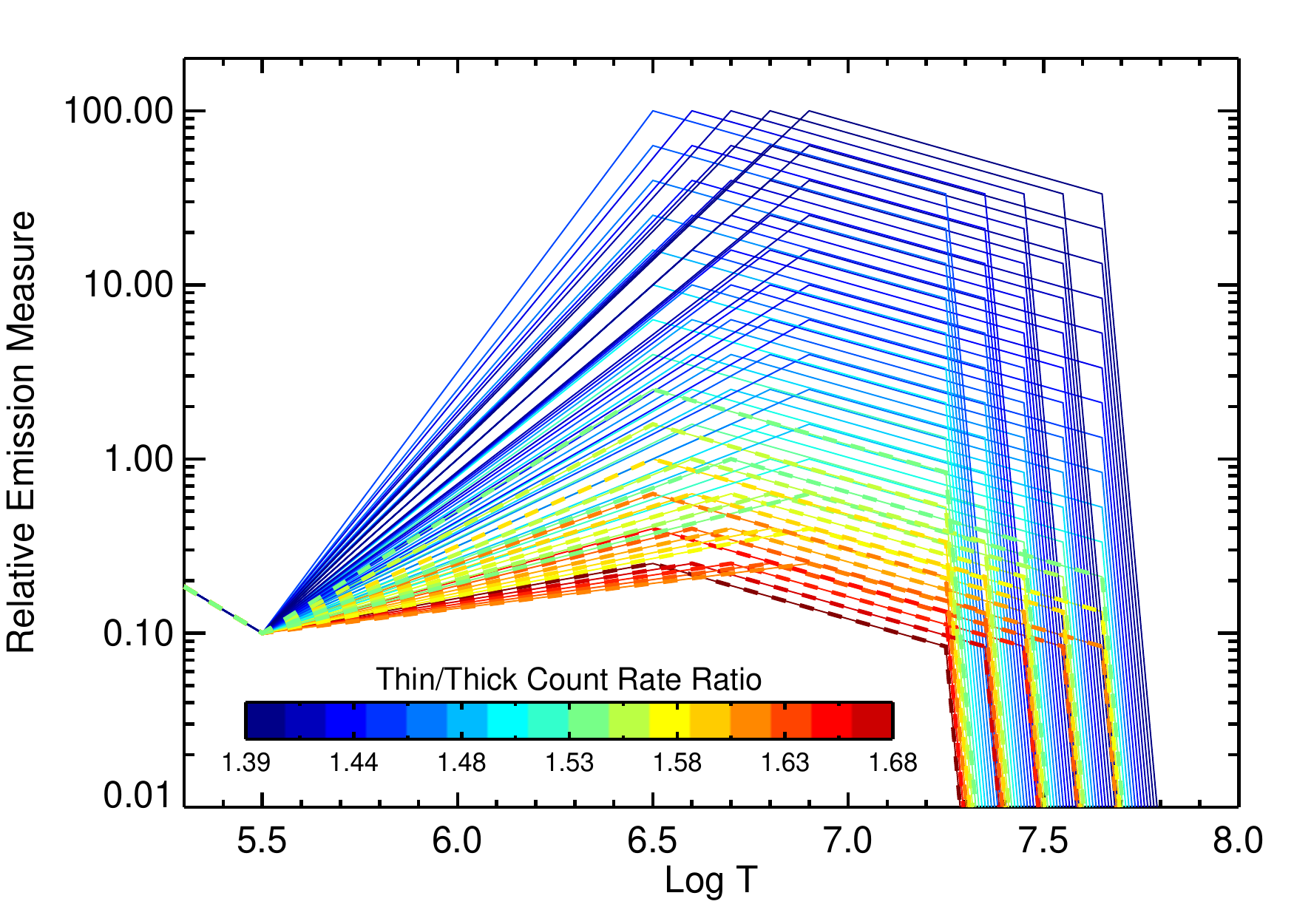} 
\caption{The idealised set of DEMs investigated for LHS~248 in this study colour-coded according to their corresponding Thin/Thick predicted count rate ratio.  Models that yielded a ratio in agreement with the observed value at the $1\sigma$ level are shown overlaid with bold dashed lines.
}
\label{f:modems}
\end{figure}

The parameters that were allowed to vary are the temperature of the peak DEM, $T_{peak}$, and the amplitude of the DEM peak relative to the DEM at $T_{min}$, specified by the scaling factor relating $\Phi(T_{peak})$ and $\log \Phi(T_{min})$.  These relations are as follows:
\begin{gather}
6.5  \leq \log T_{peak} \leq 6.9 \nonumber \\
0.4  \leq \log \Phi(T_{peak})-\log \Phi(T_{min}) \leq 3.
\label{e:dem_float}
\end{gather}
The set of model DEMs investigated are illustrated in Figure~\ref{f:modems}.  For each DEM, the synthetic spectrum was computed within the PINTofALE framework for the adopted inverse FIP abundance pattern and the ratio of thick to thin filter count rates was calculated.  DEMs for which the count rate ratio was in agreement with the $1\sigma$ uncertainty of the measured value are highlighted. 

{}

\section{Discussion}
\label{s:disc}

The off-axis pointing mode aimed at the HRC-S thick/thick Al filter boundary yielded photometric results for LHS~248 in agreement with general expectations: the count rate through the thin Al filter was found to be significantly larger than through the thick Al filter region owing to the larger effective area in the $\sim40$--170~\AA\ range in the former.  The analysis applied above suggests that this extra soft X-ray throughput is perhaps larger than might have been expected, since the observed ratio is $\sim 1.8$ while ratios from our model DEM predictions top out below 1.7.

\subsection{Isothermal Analysis}

The isothermal analysis illustrated in Figure~\ref{f:filtrat} suggests the data are compatible with a rather cool corona with a temperature $\log T< 6.4$, quite similar to the dominant temperature of the solar corona. Such a coronal temperature would be at odds with the rapid rotation of LHS~248 and its quite high X-ray luminosity in relation to its bolometric output, $L_X/L_{bol}=-3.9$.  In the presence of multi-thermal plasma, this simplistic interpretation of Figure~\ref{f:filtrat} is of course not valid. An admixture of hotter and cooler plasma could also reproduce the observed count rate ratio.  

The peak in the synthetic ratio at $\log T=5.6$--5.7 indicates that plasma at that temperature has a particularly strong influence on the relative count rates in the different filters.  This is largely due to the fall off in the short wavelength end of the spectrum such that the dominant contribution to both filter count rates shifts longward of the C~K edge.  The flatter ratio for temperatures $\log T > 6.5$ is due to the opposite effect and increasing dominance of the spectrum at shorter wavelengths where the filter transmittances are similar. The influence of the abundance pattern---inverse FIP versus a solar photospheric mixture---makes little difference to the count rate ratio.  

The small values of interstellar hydrogen column density toward the most nearby stars has similarly weak influence.  In general, the sensitivity of the filter ratio is expected to gradually diminish with increasing interstellar medium absorption as the longer wavelengths of the spectrum are increasingly attenuated.  However, the picture is complicated by strong lines of Fe IX-XI in the 170--200~\AA\ range and longward of the Al L edge where the two filter transmittances are very similar.  Hence the curves in the upper panel of Figure~\ref{f:filtrat} show quite a complicated pattern of predicted count rate ratio until column densities reach about $10^{20}$~cm$^{-2}$ at which point further increases in column yield the expected behaviour.

\subsection{How steep is the DEM?}

\begin{figure}{}
\center
\includegraphics[width=0.46\textwidth]{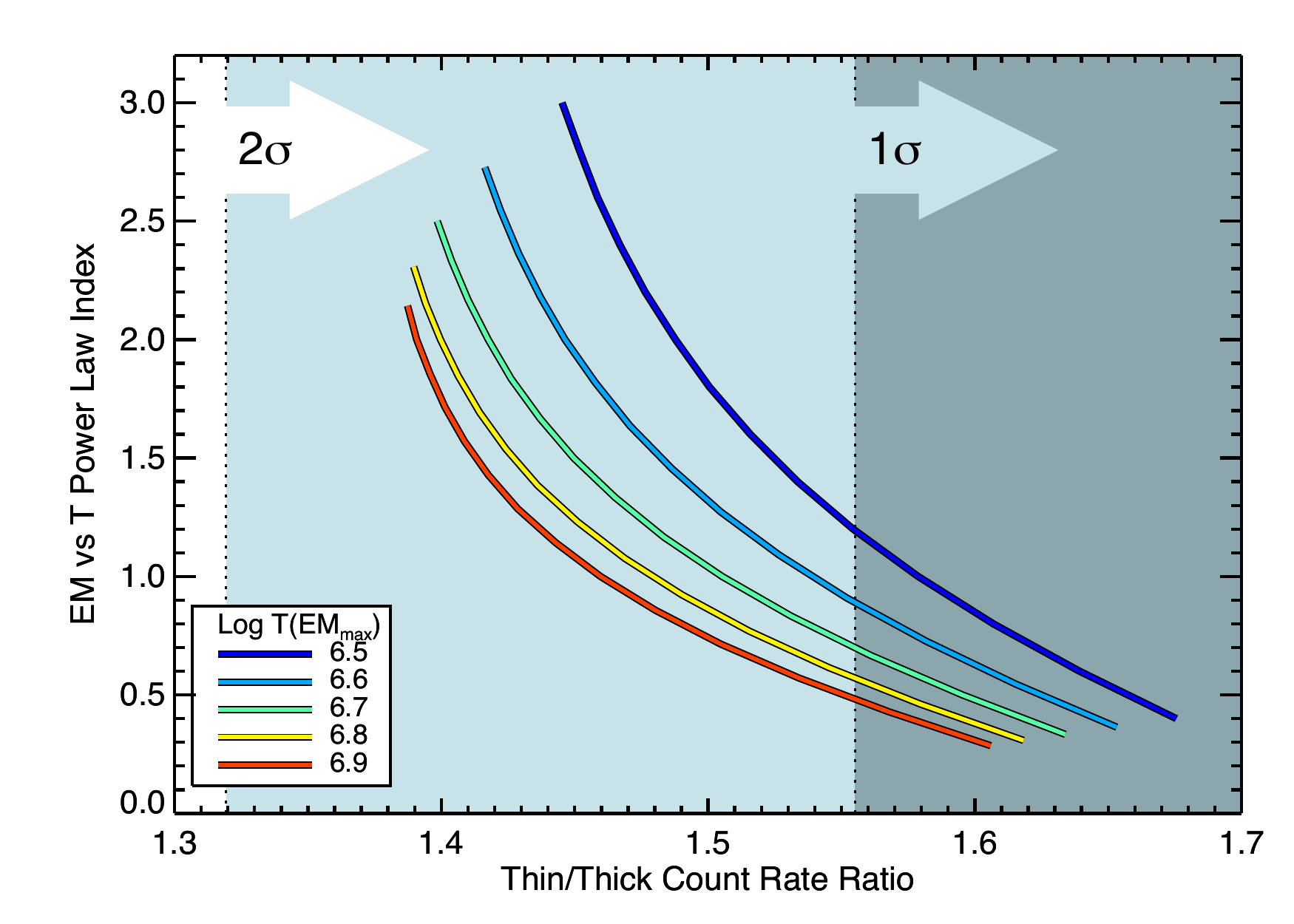} 
\caption{Loci of the EMD power law index as a function of model thin/thick filter count rate ratio. The $1\sigma$ and $2\sigma$ lower limits to the observed ratio of $1.79\pm 0.24$ for LHS~248 are indicated.
}
\label{f:emslope}
\end{figure}

The surprising result from the model differential emission measure distribution analysis is that shallow DEM slopes are favored by the LHS~248 data, in constrast to the results from the Proxima spectra.  The thin/thick count rate ratio is illustrated for the model DEMs as a function of their power law slopes from the emission measure minimum to the peak, $\alpha_{LC}$ in our model notation, in Figure~\ref{f:emslope}.  It is clear that no model is strongly excluded, since all satisfy the observations within the $2\sigma$ limit (the observation analysed here represents a pilot study; a longer exposure would have helped in providing more stringent constraints).  Figure~\ref{f:emslope} indicates that slopes of $\alpha_{LC}\leq 3/2$ are favoured.  

{}

The value of the slope of the coronal emission measure has been examined in detail dating back to earlier solar studies.  Those studies found values close to $\alpha_{LC}=3/2$ \citep[e.g.][]{Athay:66,Jordan:75,Jordan:76}, a result that can be understood in terms of the energy balance in spherically-symmetric hydrostatic equilibrium and constant cross-section loop models considering the dissipation of heating within the corona and cooling through radiation and conduction back to the chromosphere.  Extensive discussions have been provided by, e.g., \citet{Jordan:76,Craig.etal:78,Rosner.etal:78,Jordan:80,van_den_Oord.etal:97} and more recently by \citet{Jordan.etal:12}.

From a purely empirical perspective, the coronal emission measure distribution based on EUV lines observed in {\it EUVE} spectra is seen to steepen in more active stars, with values of $\alpha_{LC}\sim 2$--4 having been derived \citep[e.g.][]{Laming.etal:96,Laming.Drake:99,Sanz-Forcada.etal:03}, similar to the cores of the brightest active regions on the Sun \citep{Drake.etal:00}.  The steeper slopes are possibly the result of flare rather than steady heating \citep[e.g.][]{Gudel:97,Audard.etal:00,Kashyap.etal:02}.  Similar steeper slopes for active single and binary stars of a variety of spectral types were subsequently derived from X-ray observations with {\it Chandra} and {\it XMM-Newton} diffraction gratings \citep[e.g., beginning with][]{Drake.etal:01,Huenemoerder.etal:01,Huenemoerder.etal:03,Huenemoerder.etal:06,Telleschi.etal:05}.  We therefore {\em expect} the DEM to be somewhat more steep than $\alpha_{LC}\sim 3/2$. This expectation is of course predicated on extrapolation of the behaviour of higher mass stars to late M dwarfs.

The different chromospheric and coronal behaviour of very low mass stars compared with stars of mid-M spectral type and earlier was briefly referred to in Sect.~\ref{s:intro}.  The fractional X-ray output, $L_X/L_{bol}$, of the lowest mass stars is systematically lower than for higher masses \citep[e.g.][]{Fleming.etal:03}, exhibits a strong enhancement of the ``supersaturation'' effect---that X-ray output decreases with increasing rotation rate for the fastest rotators---and the scatter in X-ray output at a given rotation period is three times larger than for higher mass stars \citet{Cook.etal:14}. Chromospheric H$_\alpha$ fluxes are also relatively depressed \citep{Delfosse.etal:98,Mohanty.Basri:03,Fleming.etal:03}.
In contrast, the radio output of the lowest mass stars is higher than expected based on the behaviour of stars of higher mass \citep{Berger:06,Berger.etal:08,Williams.etal:14}.  

The generally favoured explanation for these changes in behaviour is a speculated change in the underlying magnetic dynamo in fully-convective stars and the form of surface magnetic field that is produced, perhaps coupled with the effects of a very low electrical conductivity in the quasi-neutral atmospheres.  \citet{Wright.Drake:16} have shown that there is no obvious change in X-ray behaviour {\em across} the fully-convective limit, with slowly rotating fully-convective stars like Proxima and Barnard's Star adhering reasonably closely to the rotation-activity relation of stars with a central radiative zone.  The results were further bolstered with a larger sample by \citet{Wright.etal:18}.   The latest spectral type of the slow rotators in the \citet{Wright.etal:18} sample is M6---comparable to the type at which the magnetic activity indicators appear to change in behaviour. Thus the existing X-ray-rotation studies do not provide any further strong clues as to the nature of the dynamo in the latest M types.

In other aspects, the characteristics of the X-ray emission of late M dwarfs is not conspicuously different to their earlier counterparts, though there are indications of a possible decline in coronal temperature.  While \citet{Wheatley.etal:17} found the X-ray emission from the M8 dwarf TRAPPIST-1 could be approximated with a two-temperature optically-thin plasma radiative loss model with temperatures of 0.15 and 0.83~keV,  \citet{Fleming.etal:03} found the temperature of the corona of the M8 dwarf VB~10 to be only 0.24~keV. 

\subsection{Missing lines in the 30--170 \AA\ range?}

Any interpretation of X-ray and EUV photometry in terms of plasma radiative loss models relies to some extent on the completeness of the model spectrum.  There are indications that plasma models are still incomplete in the 30-170~\AA\ range.  This ``missing lines'' problem was first raised in the context of the analysis of {\it EUVE} spectra: global model fits to spectra in the 70-170~\AA\ range found weaker line-to-continuum ratios than expected that were interpreted as either optical depth due to resonance scattering, or else low metal abundances, or very hot plasma that contributed only continuum to the region. The explanation is instead that flux in lines absent from the models was misinterpreted as continuum \citep[see, e.g., the discussions in][]{Drake:96,Schmitt.etal:96,Drake.etal:97}.  

Before the launch of {\it Chandra}, the 30--70~\AA\ spectral range covered by the LETGS had also been seldom observed and models were relatively poorly tested in this region. 
\citet{Testa.etal:12} used the {\it Chandra} LETGS spectrum of Procyon to evaluate atomic data relevant to the narrow-band filters of the {\it Solar Dynamics Observatory} and concluded that the {\it CHIANTI} model (Version 6) was missing flux amounting to up to a factor of 3 relative to the observed spectrum in some regions of the 50--130~\AA\ range. \citet{Testa.etal:12} noted other databases shared the same problem.  The culprit appears to be some missing transitions from ions of abundant elements such as Ne, Mg, Si, S and Ar with $n=2$ ground states, and from Fe ions with $n=3$ ground states \citep[e.g.][]{Jordan:96,Lepson.etal:05,Liang.Zhao:10,Testa.etal:12}. 

These missing lines are more prominent in cooler coronal spectra---the DEM for Procyon peaks around $\log T\sim 6.3$, for example \citep[e.g.][]{Drake.etal:95}.
Since the missing model flux is greater in the spectral region where the Thin filter has a significantly larger effective area than that Thick region, the effect would be to {\em underestimate} the Thin/Thick ratio for a given emission measure distribution.  In order to compensate for this, the DEM would need to be more shallow so as to have more cooler plasma that can mimic the missing flux. It is possible, then, that the relatively shallow DEM slopes favoured by the LHS~248 observation are at least partly an artifact of deficiencies in the radiative loss model.


\subsection{Utility of the Thin/Thick ratio for predicting the EUV flux}

\begin{figure}{}
\center
\includegraphics[width=0.46\textwidth]{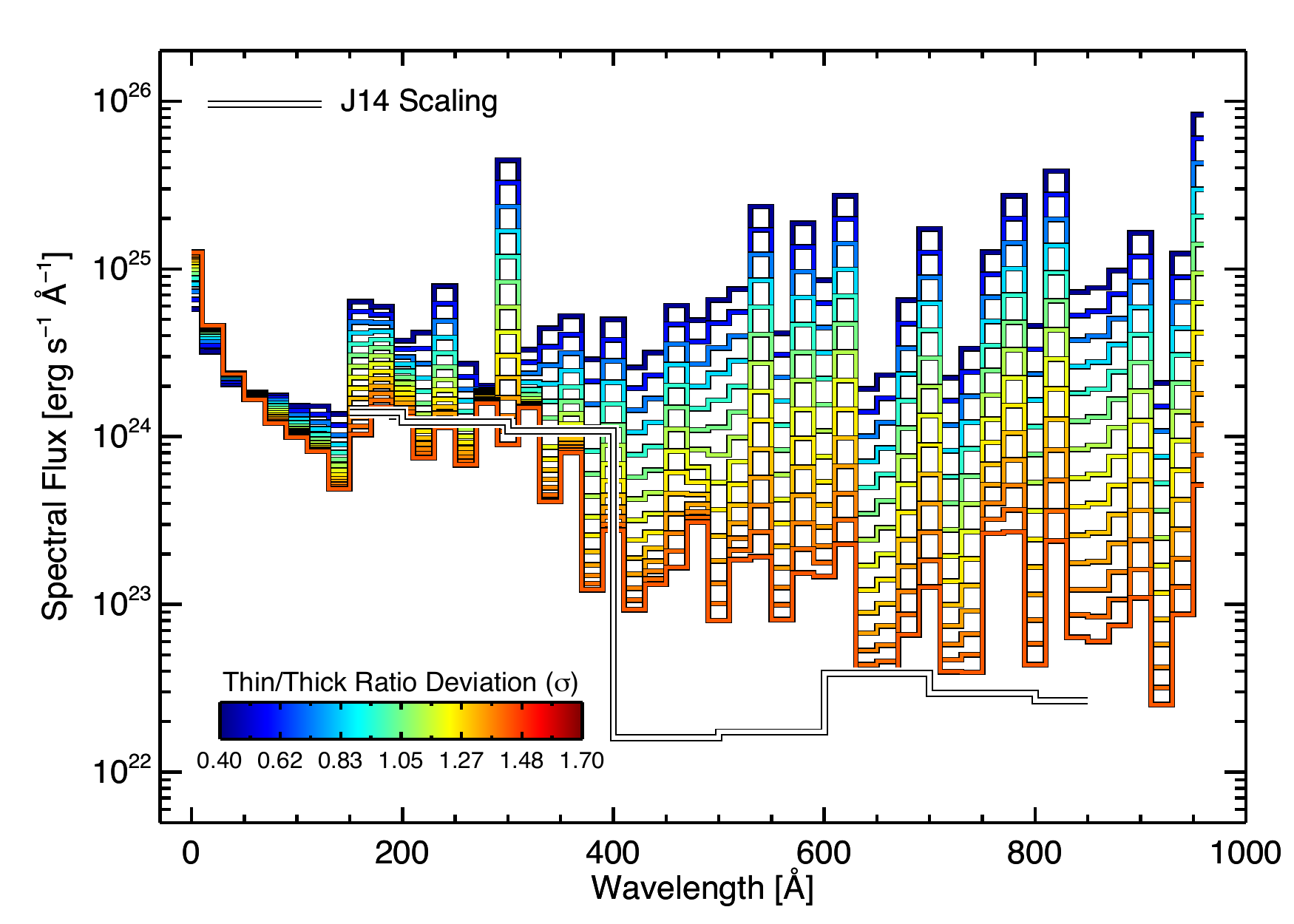}
\caption{Synthetic spectra computed from the model DEMs with $\log T_{peak}=6.5$ for inverse FIP abundances binned on 20~\AA\ intervals and colour-coded according to the deviation in their predicted thin/thick filter ratios from the observed LHS~248 value. The flux predicted by the \citet{Linsky.etal:14} empirical scaling based on the Ly$\alpha$ flux of Proxima, scaled to the radius of LHS~248 is also shown (denoted J14).
}
\label{f:modemeuv}
\end{figure}

The ultimate goal of the experiment was to design an observation that would provide information to help assess coronal emission in the 100-900~\AA\ EUV spectral range of stars that are too faint to be observed with the {\it Chandra} LETGS. To provide an illustration of the EUV leverage of our observations of LHS~248, the set of X-ray--EUV spectra predicted by the DEM models with a temperature at DEM maximum of $\log T_{peak}=6.5$ and binned on 20~\AA\ intervals (for other values of $T_{peak}$ within the same confidence range spectra are very similar) are illustrated in Figure~\ref{f:modemeuv}.  

{}

The EUV flux shortward of about 400~\AA\ is dominated by temperatures $\log T > 5.5$, a regime for which our modelling approach should be able to provide an accurate description.  Toward longer wavelengths, emission is dominated by cooler plasma and our assumptions regarding the emission measure at its minimum, $\Phi(T_{min})$, and at cooler temperatures, $\Phi(10^4\,K)$, begin to be important.  Additionally, large contributors to the flux in the 800--900~\AA\ and 450-500~\AA\ ranges are the H Ly$\alpha$ and He~I continua that are formed largely in the chromosphere and lower transition region \citep[see also, e.g.,][]{Linsky.etal:14}.  While our DEM models extend down to chromospheric temperatures, assumptions of optically-thin, collision-dominated thermal equilibrium tend to break down there and reproduction of the recombination continuum is not likely to be wholly accurate. 

Proceeding with the assumption that missing lines do not have a significant impact on the photometric filter ratio interpretation, Figure~\ref{f:modemeuv} demonstrates that the general uncertainty in the flux at 100~\AA\ within the $1\sigma$ range given by the spread in the model fluxes is only about 25\%.  Uncertainties grow, as would be expected, toward longer wavelengths.  In the 200--300~\AA\ region the $1\sigma$ range is a factor of 2--3. The flux at 300~\AA\ is dominated by the He~II Ly$\alpha$ transition at 303.78~\AA\ whose formation in the chromosphere,  as for the H I and He I continua, is likely subject to complications from photoionization and recombination and thus might not be accurately represented by optically-thin collision-dominated models in thermal equilibrium \citep[e.g.][and references therein]{Jordan.etal:93}.  The flux uncertainty grows to a factor of 4--5 in the 300-400~\AA\ range, and beyond that is about an order of magnitude. 

These considerations lead us to conclude that the observational method developed in this paper can provide a reasonably accurate extrapolation of the EUV flux of stars.  Given observations of higher S/N than attained for LHS~248, tight constraints on EUV emission in the 100-400~\AA\ range are achievable.  The efficacy of the filter ratio method is somewhat dependent on the completeness of radiative loss models; a re-examination of this issue would be strongly motivated.

Figure~\ref{f:modemeuv} also illustrates the EUV-Ly$\alpha$ scaling relations derived by \citep{Linsky.etal:14} that are in common use.  There appear to be no existing Ly$\alpha$ measurements for LHS~248, and so we have simply scaled the flux for Proxima by the  relative surface areas, $R_{248}/R_{Prox}$, to set the normalization.  The agreement with our models shortward of 400~\AA\ is good for the models with steeper DEM slopes. However, the scaling relations predict a strong discontinuity in the EUV flux at 400~\AA\ by almost a factor of 100, in the sense that the flux in the 400-600~\AA\ range appears to be strongly underpredicted.  It can be seen from our spectral models that such a discontinuity is unphysical in the sense that no superposition of plasma temperatures can conspire to give a spectrum with such a large jump in the flux at that wavelength. The {\em minimum} flux in the 400-600~\AA\ range is given by purely hot coronal emission with temperatures in excess of $10^6$~K that are adequately constrained by X-ray observations.  Plasma at temperatures $< 10^6$~K only adds to this. 

\section{Conclusions}
\label{s:con}

A specially-designed off-axis {\it Chandra} HRC-S observation of the late M dwarf LHS~248 succeeded in placing the  source on the boundary between Thin and Thick Al coating regions of the UV/ion shield filter.  Dithering the source between the two regions of the filter resulted in a modulation of the observed count rate such that the rate for the Thin filter was consistently higher than for the Thick. The count rate difference is due to the effective area of the Thin filter being larger by up to a factor of 4 over that of the Thick filter in the 44--170~\AA\ range.  

A set of model DEMs based on the DEM shape found from an analysis of an archival {\it Chandra} HETG spectrum of Proxima and also guided by results in the literature was generated.  These DEMs were used to compute model spectral that were assessed according to their ability to reproduce the observed Thin/Thick count rate ratio.  Shallow power law slopes to the DEM in the temperature range $5.5 \leq \log T \leq 6.5$ are favoured at the $1\sigma$ level, but steeper slopes are accommodated within a $2\sigma$ range.  It is possible that the shallow slopes are an artifact of incompleteness in current radiative loss models in the 30--170~\AA\ range; re-examination of this issue would be strongly motivated.

The modelling approach adopted here indicates that, in principle, {\it Chandra} HRC-S Thin/Thick photometric observations can provide an accurate---at least within a factor of 2-4---estimate of the EUV fluxes of stars in the 100-400~\AA\ range.  Constraints at longer wavelengths require more accurate knowledge of the cooler $\log T < 5.5$ emission measure.  We also find that the commonly used EUV scaling relations for M dwarf spectral types of \citet{Linsky.etal:14} have an unphysical discontinuity at 400~\AA\ and likely underestimate the flux in the 400-600~\AA\ range.


\acknowledgments
This work was funded by NASA {\it Chandra} grant GO8-19015X and NASA contract NAS8-03060 to the {\it Chandra X-ray Center}.  JJD thanks the Director, Belinda Wilkes, for continuing advice and support. 


\facility{{\it Chandra} X-ray Observatory}
\software{PINTofALE \citep{Kashyap.Drake:00}, CIAO \citep{Fruscione.etal:06}}

\bibliographystyle{yahapj}
\bibliography{jjdrake,lhs248,chandra_spie,chandra,proxima,dem,corabun,ism,verylate,photevap,cr,planetevap,euv,he304,dem}

\end{document}